\documentclass[seceq,preprint]{ptptex}

\usepackage{graphicx}
\usepackage{wrapft}
\usepackage{amsmath,amssymb}
\usepackage{type1cm}
\usepackage{bbm}

\DeclareMathOperator{\sign}{sign}
\DeclareMathOperator{\tr}{tr}

\DeclareMathOperator{\Tr}{Tr}
\DeclareMathOperator{\Pf}{Pf}

\DeclareMathOperator{\Real}{Re}
\DeclareMathOperator{\Arg}{Arg}
\newcommand{\rmd}{{\rm d}}

\newcommand\fverb{\setbox\pippobox=\hbox\bgroup\verb}
\newcommand\fverbdo{\egroup\medskip\noindent%
                        \fbox{\unhbox\pippobox}\ }
\newcommand\fverbit{\egroup\item[\fbox{\unhbox\pippobox}]}
\newbox\pippobox

\preprintnumber[3cm]{
RIKEN-TH-122\\OIQP-07-16
}

\title{
Observing Dynamical Supersymmetry Breaking with Euclidean Lattice
Simulations%
}

\author{
Issaku \textsc{Kanamori},$^{1,}$\footnote{E-mail: kanamori-i@riken.jp}
Fumihiko \textsc{Sugino}$^{2,}$\footnote{E-mail:
fumihiko\_sugino@pref.okayama.lg.jp}
and Hiroshi \textsc{Suzuki}$^{3,}$\footnote{E-mail: hsuzuki@riken.jp}
}

\inst{
$^{1,3}$Theoretical Physics Laboratory, RIKEN, Wako 351-0198, Japan\\
$^2$Okayama Institute for Quantum Physics, Okayama 700-0015, Japan
}

\recdate{November 15, 2007}

\abst{
A strict positivity of the ground-state energy is a necessary and sufficient
condition for spontaneous supersymmetry breaking. This ground-state energy
may be directly determined from the expectation value of the Hamiltonian in
the functional integral, defined with an \emph{antiperiodic\/} temporal
boundary condition for all fermionic variables. We propose to use this fact to
observe the dynamical spontaneous supersymmetry breaking in Euclidean lattice
simulations. If a lattice formulation possesses a manifestly preserved
fermionic symmetry, there exists a natural choice of a Hamiltonian operator
that is consistent with a topological nature of the Witten index. We
numerically confirm the validity of our idea in models of supersymmetric
quantum mechanics. We further examine the possibility of dynamical
supersymmetry breaking in the two-dimensional $\mathcal{N}=(2,2)$ super
Yang-Mills theory with the gauge group~$SU(2)$, for which the Witten index is
unknown.
Although statistical errors are still large, we do not observe positive
ground-state energy, at least within one standard deviation. This prompts us to
draw a different conclusion from a recent conjectural claim that supersymmetry
is dynamically broken in this system.}

\begin{document}

\maketitle

\section{Introduction}
The possibility of the spontaneous breaking of supersymmetry (assuming that it
is not broken at the tree level) is a highly dynamical issue and its precise
study requires a nonperturbative framework. Generally, the Witten
index~\cite{Witten:1982df} provides an important clue. One can infer that
dynamical supersymmetry breaking does not occur in a wide class of
supersymmetric models where the Witten index can be computed to be nonzero.
However, the Witten index is not a panacea. There are still many interesting
models for which it is very difficult to determine the Witten index and, in
some cases, the index itself would be ill-defined because of a gapless
continuous spectrum.\cite{deWit:1988ct,Sethi:1997pa}

On the other hand, it is well known that a strict positivity of the
ground-state (or vacuum) energy is a necessary and sufficient condition
for spontaneous supersymmetry breaking.\cite{Witten:1981nf} \ In principle,
therefore, one can judge whether supersymmetry breaking occurs by
computing the ground-state energy.

In this paper, in light of recent developments on the lattice formulation of
supersymmetric
theories,\cite{Kaplan:2003uh,Feo:2004kx,Giedt:2006pd,Giedt:2007hz} we propose
to observe dynamical supersymmetry breaking with Euclidean lattice formulations
employing the above idea. This work was originally motivated by a recent
paper by Hori and Tong~\cite{Hori:2006dk} in which they conjectured dynamical
supersymmetry breaking in the two-dimensional $\mathcal{N}=(2,2)$
super Yang-Mills theory with the gauge group~$SU(N_c)$. Lattice formulations of
this two-dimensional theory are the simplest among recent lattice formulations
of extended supersymmetric gauge theories.\cite{Kaplan:2002wv,
Cohen:2003xe,Sugino:2003yb,Sugino:2004qd,Catterall:2004np,Suzuki:2005dx,
D'Adda:2005zk,Sugino:2006uf,Catterall:2006jw,Catterall:2006is,Suzuki:2007jt,
Fukaya:2007ci} \ Therefore, it is highly natural, if it is possible, to examine
supersymmetry breaking in the two-dimensional $\mathcal{N}=(2,2)$ super
Yang-Mills theory with lattice formulation.\footnote{While preparing this
paper, we discovered a preprint\cite{Matsuura:2007ec} in which this problem
is addressed on the basis of the lattice formulation in
Ref.~\citen{Cohen:2003xe}.}
This is what we do in this study. To our knowledge, this is the first instance
in which the dynamical supersymmetry breaking in gauge field theory is
investigated numerically, although there exists a closely related and
thought-provoking observation in Ref.~\citen{Catterall:2006is}. (There are a
number of numerical works related to this issue in one-dimensional
supersymmetric models\cite{Catterall:2000rv,Wosiek:2002nm,Campostrini:2002mr,
Campostrini:2004bs,Giedt:2004vb,Kotanski:2006wp,Bergner:2007pu,Hanada:2007ti,
Catterall:2007fp,Anagnostopoulos:2007fw} and two-dimensional Wess-Zumino
models.\cite{Ranft:1983ag,Schiller:1986zx,Beccaria:1998vi,Catterall:2003wd,
Beccaria:2004pa}) 
Although statistical errors in our Monte Carlo study using a formulation
in~Ref.~\citen{Sugino:2004qd} are still large, we do not observe positive
ground-state energy, at least within one standard deviation. This observation
prompts us to draw a different conclusion from the conjectural claim in
Ref.~\citen{Hori:2006dk}.
This is the content of \S4. In \S2, we present our basic idea concerning the
determination of the ground-state energy in the Euclidean functional integral
formalism. Then, in \S3, we illustrate how our method works by applying it to
supersymmetric quantum mechanical
models.\cite{Witten:1981nf} \ Section~5 is devoted to the discussion.

In what follows, the boundary condition of fermionic variables for the temporal
direction, whether it is periodic (PBC) or antiperiodic (aPBC), is crucial.
For all bosonic variables and for all variables with respect to the
spatial directions, we will assume the periodic boundary conditions.
Unless noted otherwise, the term ``boundary condition'' will always refer to
the boundary condition of fermionic variables for the temporal direction.

\section{Basic idea}
What we want to determine is the ground-state energy~$E_0$ of supersymmetric
theories. If $E_0>0$, supersymmetry is spontaneously broken and it is not if
$E_0=0$.\cite{Witten:1981nf} \ With the Euclidean functional integral
formalism, one could determine the ground-state energy from the expectation
value of Hamiltonian~$H$,
\begin{equation}
   \langle H\rangle_{\text{PBC}}
   =\frac{\int_{\text{PBC}}\rmd\mu\,He^{-S}}{\int_{\text{PBC}}\rmd\mu\,e^{-S}},
\label{twoxone}
\end{equation}
where $S$ is the Euclidean action and $\rmd\mu$ symbolically denotes a measure
for the functional integration. We assumed the periodic boundary condition
(PBC) for fermionic variables because this boundary condition is
consistent with supersymmetry. One would then be able to obtain the
ground-state energy~$E_0$ by taking the large imaginary-time limit
$\beta\to\infty$, where $\beta$ is the temporal size of the
system,\footnote{$\beta$ is not to be confused with the conventional gauge
coupling constant in lattice gauge theory.} because in this limit, only the
contribution of the ground-state(s) survives in~Eq.~(\ref{twoxone}).

However, this naive idea is wrong. First, we must note that the functional
integral in the denominator of Eq.~(\ref{twoxone}) is proportional to the
Witten index,\cite{Cecotti:1981fu,Fujikawa:1982nt}\footnote{Our discussion in
this section is based on the assumption that the expressions appearing in
Eq.~(\ref{twoxtwo}) are meaningful. For this, we may assume that the spectrum
of~$H$ is discrete so that the Witten index is unambiguously defined. Our basic
formula~(\ref{twoxseven}) for the ground-state energy itself, however, might
also be applicable to systems in which this assumption fails.}
\begin{equation}
   \mathcal{Z}_{\text{PBC}}
   \equiv\mathcal{N}_{\text{PBC}}\int_{\text{PBC}}d\mu\,e^{-S}
   =\Tr(-1)^Fe^{-\beta H}=\Tr(-1)^F,
\label{twoxtwo}
\end{equation}
where $F$ is the fermion number operator and $\mathcal{N}_{\text{PBC}}$ is a
proportionality constant that depends on the choice of the integration
measure~$d\mu$. The constant $\mathcal{N}_{\text{PBC}}$ may depend on
ultraviolet and infrared cutoffs (the number of lattice points for lattice
regularization) and possibly on the boundary condition. Second, the numerator
of Eq.~(\ref{twoxone}) is proportional to the derivative of the Witten index
with respect to $\beta$, which is \emph{always\/} zero:
\begin{equation}
   \mathcal{N}_{\text{PBC}}\int_{\text{PBC}}d\mu\,He^{-S}
   =\Tr(-1)^FHe^{-\beta H}
   =-\frac{\partial}{\partial \beta}\Tr(-1)^Fe^{-\beta H}
   =0.
\label{twoxthree}
\end{equation}
This independence of the Witten index from a parameter of the theory, $\beta$,
is a consequence of the supersymmetry algebra.\cite{Witten:1982df} \ Thus we
have
\begin{equation}
   \langle H\rangle_{\text{PBC}}=\frac{0}{\Tr(-1)^F}.
\label{twoxfour}
\end{equation}
We finally recall that the Witten index vanishes when supersymmetry is
spontaneously broken. Therefore, we see that $\langle H\rangle_{\text{PBC}}$ is
indefinite when supersymmetry is broken. Otherwise, it is zero or indefinite
depending on whether or not the Witten index is nonzero. Note also that a
similar remark is valid for the expectation value of generic operators when
the periodic boundary condition is imposed; when the Witten index vanishes,
$\mathcal{Z}_{\text{PBC}}$ cannot be used as a normalization factor for
expectation values. With the periodic boundary condition, therefore, the
expectation values normalized by the partition function can be ill-defined, and
in such a case, we must consider ``denominator-free'' expectation values, such
as Eq.~(\ref{twoxthree}). In any case, the expectation value
$\langle H\rangle_{\text{PBC}}$ does not provide useful direct information on the
ground-state energy or on supersymmetry breaking.

In Eq.~(\ref{twoxthree}), what prevents us from obtaining the ground-state
energy is the factor $(-1)^F$, which is the heart of the Witten index. If this
factor can be removed, the above idea of using the expectation value of the
Hamiltonian would be valid. As is well known (see, for example,
Refs.~\citen{Girardello:1980vv,Fujikawa:1982nt}), such a removal can easily be
achieved. What we must do is simply to change the boundary condition of
fermionic variables for the temporal direction from periodic to anti\-periodic
(aPBC). This defines the thermal partition function with the inverse
temperature~$\beta$, instead of the Witten index,
\begin{equation}
   \mathcal{Z}_{\text{aPBC}}
   \equiv\mathcal{N}_{\text{aPBC}}\int_{\text{aPBC}}d\mu\,e^{-S}
   =\Tr e^{-\beta H},
\label{twoxfive}
\end{equation}
which should be positive definite, and, as its derivative with respect to
$\beta$,
\begin{equation}
   \mathcal{N}_{\text{aPBC}}\int_{\text{aPBC}}d\mu\,He^{-S}
   =\Tr He^{-\beta H}.
\label{twoxsix}
\end{equation}
Therefore, taking the long-time limit (or the low-temperature limit) of the
ratio of these two quantities, we have
\begin{equation}
   \lim_{\beta\to\infty}\langle H\rangle_{\text{aPBC}}
   =\lim_{\beta\to\infty}
   \frac{\int_{\text{aPBC}}d\mu\,He^{-S}}
   {\int_{\text{aPBC}}d\mu\,e^{-S}}
   =\lim_{\beta\to\infty}\frac{\Tr He^{-\beta H}}{\Tr e^{-\beta H}}
   =E_0,
\label{twoxseven}
\end{equation}
and ground-state energy~$E_0$ is obtained. This is our basic formula.

Before taking the $\beta\to\infty$ limit, Eq.~(\ref{twoxseven}) is merely the
expectation value in the thermal equilibrium with finite
temperature~$1/\beta$, and supersymmetry is explicitly broken by the
temperature. In this aspect, it is interesting to note an analogy to a
conventional way of detecting the spontaneous breaking of an ordinary symmetry,
for example, the $Z_2$ symmetry of the Ising spin. In this case, one breaks the
symmetry by applying an external magnetic field that is conjugate to the order
parameter of symmetry breaking, that is the magnetization. One then observes
(in the thermodynamic limit) how a trace of the breaking remains after the
applied field is turned off.

For supersymmetry, the order parameter is a positivity of the ground-state
energy and the conjugate variable to the energy is the temperature. In
Eq.~(\ref{twoxseven}), we break supersymmetry by placing the system in
thermal equilibrium. We then observe how the effect of the temperature
remains in the zero-temperature (or the large imaginary-time) limit
$\beta\to\infty$, for which one naively would expect that the effect simply
disappears. If the effect remains, we judge that spontaneous supersymmetry
breaking occurs. Recall that the expectation
value~$\langle H\rangle_{\text{PBC}}$ with the periodic boundary condition is
always zero or indefinite. Thus, if a well-defined $E_0>0$ is found through
Eq.~(\ref{twoxseven}), it is the effect of the boundary condition surviving
even in the long-time (or the zero-temperature)
limit.\footnote{Note, however, that spontaneous supersymmetry breaking
differs from spontaneous breaking of ordinary symmetries in that
it can occur even in a system with finite volume.\cite{Witten:1981nf}}
Physically, this survival of the effect of the boundary condition can be
understood in terms of the appearance of a massless (or zero energy for quantum
mechanics) Nambu-Goldstone fermion associated with spontaneous supersymmetry
breaking.

Our basic formula~(\ref{twoxseven}) is very simple. However, to embody it in
Euclidean lattice formulation, there are still several issues to be clarified.
First of all, the above argument assumes that regularization to define the
functional integral does not break supersymmetry. Otherwise, one would not be
able to distinguish spontaneous supersymmetry breaking from a possible
explicit breaking due to regularization. As is well recognized, generally,
a regularization based on a spacetime lattice is irreconcilable with
supersymmetry. For theories with the extended supersymmetry, it is nevertheless
sometimes possible to set up a lattice regularization that preserves the
invariance under some supersymmetry
transformations.\cite{Kaplan:2003uh,Feo:2004kx,Giedt:2006pd,Giedt:2007hz}
\ Then, if the spacetime dimension is low enough, one may expect that the
invariance under a full set of supersymmetry transformations is restored in the
continuum limit. In what follows, we assume this sort of lattice
regularization. In particular, for the study of the two-dimensional
$\mathcal{N}=(2,2)$ super Yang-Mills theory, we adopt the formulation in
Ref.~\citen{Sugino:2004qd} in which a fermionic symmetry~$Q$ that is a part of
the supersymmetry is manifestly preserved. We will briefly review this
formulation in~\S4.

Secondly, closely related to the above point, we must properly choose a
possible additive constant in the Hamiltonian~$H$. In other words, we must
correctly choose the origin of the energy. This point is, of course, crucial
for judging spontaneous supersymmetry breaking from the positivity
of~$E_0$. Note also that, when the Witten index is nonzero, supersymmetric
invariant state(s) must have a precisely zero energy eigenvalue $E_0=0$ for
relation~(\ref{twoxthree}) to hold. That is, Eq.~(\ref{twoxthree}) is not
invariant under an arbitrary shift of the origin of the energy $H\to H+c$, when
the Witten index is nonzero.

Of course, a natural prescription for defining the Hamiltonian is to use the
supersymmetry algebra. One may first define supercharge operators $\mathcal{Q}$
and $\mathcal{Q}^\dagger$ (with some regularization) and define a (regularized)
Hamiltonian operator~$H$ by the anti-commutation relation
$H=\{\mathcal{Q},\mathcal{Q}^\dagger\}/2$ without any additive constant. This is
precisely the idea behind the Hamiltonian formulation of supersymmetric
theories\cite{%
Elitzur:1982vh,Sakai:1983dg,Elitzur:1983nj,Claudson:1984th,
Wosiek:2002nm,Campostrini:2002mr,Campostrini:2004bs,Kotanski:2006wp,
Ranft:1983ag,Schiller:1986zx,Beccaria:2004pa} with which a possible additive
constant in the Hamiltonian~$H$ is automatically fixed.\footnote{On the other
hand, from the viewpoint of the feasibility of numerical simulations that
preserve the gauge symmetry, the Euclidean lattice formulation appears
advantageous.}

For the following reason, however, this issue of a ``correct'' Hamiltonian is
somewhat delicate in the functional integral formulation based on the
Lagrangian.

Suppose that the (for simplicity, off-shell) supersymmetry algebra is realized
by the transformation law for variables appearing in the continuum Lagrangian.
This implies that there exists a fermionic \emph{transformation\/} $Q$ such
that $\{Q,\overline Q\}=2i\partial_0$, where $\overline Q$ is the conjugate
fermionic transformation of $Q$ and $\partial_0$ is the time derivative. One
would then expect, from this algebra, that the relation
$iQ\overline{\mathcal{Q}}=2H$ holds,\footnote{We take normalization of the
Noether charge such that the Poisson bracket
$\{\overline{\mathcal{Q}},\cdot\}_{\text{P}}$ generates the $\overline Q$
transformation. After the quantization, the relation would be read as
$H=\{\mathcal{Q},\overline{\mathcal{Q}}\}/2$ which is consistent with the
positivity of the Hamiltonian.} where $\overline{\mathcal{Q}}$ is the
\emph{Noether charge}\footnote{In field theories, when supersymmetry is
spontaneously broken, the Noether charge (supercharge) itself would be
ill-defined owing to a massless singularity associated with the Nambu-Goldstone
fermion. In the field theory case discussed in~\S4, we use the Noether current
instead.}
associated with~$\overline Q$ and $H$ is the Hamiltonian obtained from the
Lagrangian by the Legendre transformation. If this relation holds, this $H$
could be used in the functional integral as a Hamiltonian operator that is
consistent with the supersymmetry algebra.

In reality, however, the relation holds only \emph{up to equations of motion}.
Generally, one ends up with
\begin{equation}
   \frac{i}{2}Q\overline{\mathcal{Q}}
   =H+(\text{terms being proportional to equations of motion}).
\label{twoxeight}
\end{equation}
That is, a Hamiltonian suggested from the algebra can differ from the original
one obtained through the Legendre transformation from the Lagrangian.
This occurs very commonly, and we will encounter such a situation even in the
simplest supersymmetric system in the next section. The additional terms, which
would be negligible in the classical level, cannot be neglected in general
within the functional integral because those terms may give rise to contact
terms at a coincident point, i.e., ultraviolet-divergent constants. We thus
have two (among possibly many) options for a Hamiltonian operator in quantum
theory: one is the original Hamiltonian obtained from the Lagrangian and the
other is $iQ\overline{\mathcal{Q}}/2$. How can we be sure that we are using a
Hamiltonian with a correctly chosen additive constant before we measure the
ground-state energy? Clearly, we need some guiding principle.

We have no general answer to the above question. See also \S5. However, if the
lattice formulation one adopts possesses at least one exactly preserved
fermionic symmetry, for example, the above~$Q$, there exists a natural
prescription for a choice of the Hamiltonian. It is the left-hand side
of~Eq.~(\ref{twoxeight}), $H\equiv iQ\overline{\mathcal{Q}}/2$. This choice of
the Hamiltonian in quantum theory corresponds to ``renormalizing'' additional
terms in the right-hand side of Eq.~(\ref{twoxeight}) into $H$. This definition
is natural because the structure $H=iQ\overline{\mathcal{Q}}/2$ is suggested
from the supersymmetry algebra. Moreover, this choice has the correct origin of
the energy in the sense that it is consistent with the topological property of
the Witten index, Eq.~(\ref{twoxthree}). That is,
\begin{equation}
   \mathcal{N}_{\text{PBC}}\int_{\text{PBC}}d\mu\,He^{-S}
   =\mathcal{N}_{\text{PBC}}
   \int_{\text{PBC}}d\mu\,\frac{i}{2}Q\overline{\mathcal{Q}}\,e^{-S}
   =\mathcal{N}_{\text{PBC}}
   \int_{\text{PBC}}d\mu\,Q\left(\frac{i}{2}\overline{\mathcal{Q}}e^{-S}
   \right)
   =0,
\label{twoxnine}
\end{equation}
where we have used the $Q$-invariance of the action and of the integration
measure.\footnote{Strictly speaking, to show this relation, we must assume that
the integral $\int_{\text{PBC}}d\mu\,\overline{\mathcal{Q}}\,e^{-S}$ is
finite.} As already noted, when the Witten index is nonzero, this property
fixes the origin of the energy uniquely. For these reasons, we consider that
the definition $H\equiv iQ\overline{\mathcal{Q}}/2$ is natural. Of course, for
cases of interest, we do not know a priori whether the Witten index is nonzero
or not, and if it is zero, the above argument for the structure
$H=iQ\overline{\mathcal{Q}}/2$ based on relation~(\ref{twoxthree}) is
groundless (a shift of the origin $H\to H+c$ does not influence
Eq.~(\ref{twoxthree}) if
$\mathcal{N}_{\text{PBC}}\int_{\text{PBC}}d\mu\,e^{-S}=0$).
Nevertheless, we adopt this $Q$-exactness of the Hamiltonian as a working
hypothesis in what follows because the definition of a Hamiltonian operator
should be independent of whether or not the supersymmetry is spontaneously
broken.

\section{Supersymmetric quantum mechanics}
In this section, we examine our method by applying it to a Euclidean lattice
formulation of the supersymmetric quantum mechanics.\cite{Witten:1981nf} \ We
find that this example provides a good illustration of our method.

The Lagrangian of the supersymmetric quantum mechanics is given by
\begin{equation}
   L=\frac{1}{2}(\partial\phi)^2-\frac{1}{2}(W')^2
   +\overline\psi(i\partial-W'')\psi+\frac{1}{2}F^2,
\label{threexone}
\end{equation}
where all variables are functions of the ``time coordinate'' $x$ and $\partial$
is the derivative with respect to~$x$, $\partial\equiv\partial/(\partial x)$.
$\phi$ and $F$ are bosonic variables and $\overline\psi$ and $\psi$ are
fermionic. The superpotential $W=W(\phi)$ is a function of $\phi$ and the prime
denotes the derivative with respect to~$\phi$. With the periodic boundary
condition for all variables, the action $S=\int d x\,L$ is invariant under
the following $\mathcal{N}=2$ supersymmetry transformations:
\begin{align}
   &Q\phi=\psi,&& Q\psi=0,
\\
   &Q\overline\psi=F+i\partial\phi-W',&&
   QF=-i\partial\psi+W''\psi,
\end{align}
and
\begin{align}
   &\overline Q\phi=\overline\psi,&& \overline Q\,\overline\psi=0,
\\
   &\overline Q\psi=-F+i\partial\phi+W',&&
   \overline QF=i\partial\overline\psi+W''\overline\psi.
\end{align}
One can confirm that the transformations form the supersymmetry algebra
\begin{equation}
   Q^2=\overline Q^2=0,\qquad\{Q,\overline Q\}=2i\partial
\label{threexsix}
\end{equation}
off-shell, i.e., without using any equations of motion. In this system, it is
well known\cite{Witten:1981nf} that supersymmetry is spontaneously broken if
and only if the number of zeros of the function~$W'(\phi)$ is even, or
equivalently, $W(-\infty)$ and $W(+\infty)$ have opposite signs.
(We assumed that $|W(\pm\infty)|=+\infty$.)

A crucial fact for us is that the classical action can be expressed as the
$Q$-exact form\footnote{For this, we must note that
$\int d x\,iW'\partial\phi=\int d x\,i\partial W=0$.}
\begin{equation}
   S=\int d x\,L
   =Q\int d x\,\frac{1}{2}\overline\psi(F-i\partial\phi+W').
\end{equation}
Then the invariance of $S$ under $Q$ and~$\overline Q$ can be easily seen by
using the supersymmetry algebra~(\ref{threexsix}).\footnote{$S$ can also be
written as $S=Q\overline Q\int d x\,\frac{1}{2}(\overline\psi\psi+2W)$.}

The Hamiltonian corresponding to the Lagrangian~(\ref{threexone}) is given by
\begin{equation}
   H=\frac{1}{2}(\partial\phi)^2+\frac{1}{2}(W')^2
   +\overline\psi W''\psi-\frac{1}{2}F^2
\end{equation}
and, as we noted in Eq.~(\ref{twoxeight}), we have
\begin{equation}
   \frac{i}{2}Q\overline{\mathcal{Q}}
   =H+\frac{1}{2}F(F-i\partial\phi-W')
   +\frac{1}{2}\overline\psi(i\partial-W'')\psi,
\label{threexnine}
\end{equation}
where $\overline{\mathcal{Q}}$ is the Noether charge associated with the
$\overline Q$~invariance:
\begin{equation}
   \overline{\mathcal{Q}}=-\overline\psi(\partial\phi-iW').
\end{equation}
Since the last two terms in Eq.~(\ref{threexnine}) vanish under classical
equations of motion, $F=0$ and~$(i\partial-W'')\psi=0$,
relation~(\ref{threexnine}) is consistent with the supersymmetry algebra, at
least classically.

After the Wick rotation, $x\to-ix$ and $L\to-L$, we have the Euclidean
action
\begin{align}
   S&=\int d x\,\left\{\frac{1}{2}(\partial\phi)^2+\frac{1}{2}(W')^2
   +\overline\psi(\partial+W'')\psi-\frac{1}{2}F^2\right\}
\\
   &=-Q\int d x\,\frac{1}{2}\overline\psi(F+\partial\phi+W'),
\label{threextwelve}
\end{align}
and for the Hamiltonian~$H$,
\begin{equation}
   \frac{1}{2}Q\left\{\overline\psi(\partial\phi-W')\right\}
   =H+\frac{1}{2}F(F+\partial\phi-W')
   -\frac{1}{2}\overline\psi(\partial+W'')\psi.
\label{threexthirteen}
\end{equation}

So far everything has been for the continuum. We now construct a lattice
formulation of the above system on a finite-size lattice,
\begin{equation}
   \Lambda=\left\{x\in a\mathbb{Z}\mid0\leq x<\beta\right\},
\end{equation}
where $a$ denotes the lattice spacing. First, we fix the lattice transcription
of the time derivative~$\partial$. As a possible choice, we adopt the forward
difference
\begin{equation}
   \partial f(x)\equiv f(x+a)-f(x),
\end{equation}
which does not lead to species doubling. The lattice counterparts of the $Q$
transformation are then defined as
\begin{align}
   &Q\phi(x)=\psi(x),&& Q\psi(x)=0,
\label{threexsixteen}
\\
   &Q\overline\psi(x)=F(x)-\partial\phi(x)-W'(\phi(x)),&&
   QF(x)=\partial\psi(x)+W''(\phi(x))\psi(x).
\label{threexseventeen}
\end{align}
Finally, the lattice action is defined with an expression analogous to
Eq.~(\ref{threextwelve}):
\begin{align}
   S&\equiv-Q
   \sum_{x\in\Lambda}\frac{1}{2}
   \left\{\overline\psi(x)\left(F(x)+\partial\phi(x)+W'(\phi(x))\right)\right\}
\\
   &=\sum_{x\in\Lambda}\biggl\{
   \frac{1}{2}\partial\phi(x)\partial\phi(x)
   +\frac{1}{2}(W'(\phi(x)))^2
   +\overline\psi(x)\left(\partial+W''(\phi(x))\right)\psi(x)
\nonumber\\
   &\qquad\qquad\qquad\qquad\qquad\qquad\qquad\qquad\qquad{}
   -\frac{1}{2}F(x)^2+W'(\phi(x))\partial\phi(x)
   \biggr\},
\label{threexnineteen}
\end{align}
where
\begin{equation}
   \psi(x=\beta)=\begin{cases}
   +\psi(x=0), & \text{for the periodic boundary condition}, \\
   -\psi(x=0), & \text{for the antiperiodic boundary condition}.
\end{cases}
\end{equation}
Note that all lattice variables are dimensionless. In the above lattice
formulation with the periodic boundary condition, the $Q$-symmetry is
manifestly preserved because the $Q$-transformations in
Eqs.~(\ref{threexsixteen}) and~(\ref{threexseventeen}) are nilpotent, $Q^2=0$.
The invariance under $\overline Q$ is, however, broken because it is impossible
to define a corresponding $\overline Q$ transformation on lattice variables
such that the algebra~$\{Q,\overline Q\}=-2\partial$ still holds. In fact, this
lattice action is basically identical to the one described
in~Refs.~\citen{Catterall:2000rv,Giedt:2004vb,Bergner:2007pu,Beccaria:1998vi,
Catterall:2003wd,Giedt:2004qs}. (See also Ref.~\citen{Hanada:2007ti}.) In some
of these references, it has been shown that the $\overline Q$-symmetry is
restored in the continuum limit.

As a Hamiltonian in this lattice formulation, following the discussion in
the previous section and in view of Eq.~(\ref{threexthirteen}), we use
$H(x)\equiv iQ\overline{\mathcal{Q}}(x)/2$, where\footnote{We supplemented a
factor of $1/a$ to adjust the physical mass dimension; recall that all lattice
variables as well as the $Q$~transformation are dimensionless.}
\begin{equation}
   \overline{\mathcal{Q}}(x)\equiv-\frac{1}{a}
   \overline\psi(x)\left(i\partial\phi(x)-iW'(\phi(x))\right)
\end{equation}
is a lattice analogue of the Noether charge. The explicit form is
\begin{align}
   H(x)&=-\frac{1}{2a}\partial\phi(x)\partial\phi(x)
   +\frac{1}{2a}(W'(\phi(x)))^2
   -\frac{1}{2a}\overline\psi(x)\left(\partial-W''(\phi(x))\right)\psi(x)
\nonumber\\
   &\qquad\qquad\qquad\qquad\qquad\qquad\qquad\qquad\qquad{}
   +\frac{1}{2a}F(x)\left(\partial\phi(x)-W'(\phi(x))\right).
\label{threextwentytwo}
\end{align}
The naive continuum limit of~$H(x)$ differs from the (imaginary-time)
Hamiltonian in the continuum theory by terms vanishing under the classical
equations of motion.\footnote{It is interesting to note that the expectation
value (with the antiperiodic boundary condition) of this difference vanishes:
$\left\langle\frac{1}{2}F(F+\partial\phi-W')
-\frac{1}{2}\overline\psi(\partial+W'')\psi\right\rangle_{\text{aPBC}}=0$.}
As discussed in the previous section, this lattice Hamiltonian has a correct
zero-point energy in the sense that
\begin{equation}
   \int_{\text{PBC}}\prod_{x\in\Lambda}d\phi(x)\,d F(x)\,
   d\psi(x)\,d\overline\psi(x)\,He^{-S}=0,
\label{threextwentythree}
\end{equation}
which follows from the $Q$-exactness of $H$ and the $Q$-invariance of the
action~$S$ and of the integration measure.\footnote{Assuming that the integral
$\int_{\text{PBC}}\prod_{x\in\Lambda}d\phi(x)\,d F(x)\,d\psi(x)\,
d\overline\psi(x)\,\overline{\mathcal{Q}}\,e^{-S}$ is finite. This is
certainly true if $|W'(\pm\infty)|=+\infty$.} (Recall Eq.~(\ref{twoxnine}).)
Thus this choice of the Hamiltonian is consistent with the topological nature
of the Witten index, Eq.~(\ref{twoxthree}), with finite ultraviolet and
infrared cutoffs.

The numerical study of the present lattice model is not so difficult, because
it is possible to obtain a closed expression of the fermion determinant in terms
of~$\phi$. That is,
\begin{equation}
   \det\left\{-\partial-W''(\phi)\right\}
   =\prod_{x\in\Lambda}\left\{1-W''(\phi(x))\right\}\mp1,
\label{threextwentyfour}
\end{equation}
where the upper sign corresponds to the periodic boundary condition and the
lower corresponds to the antiperiodic boundary condition. Thus, after
integrating over fermionic variables and the auxiliary variable~$F(x)$, we have
the effective action
\begin{align}
   S_{\text{eff}}[\phi]
   &=\sum_{x\in\Lambda}\left\{
   \frac{1}{2}\partial\phi(x)\partial\phi(x)
   +\frac{1}{2}(W'(\phi(x)))^2
   +W'(\phi(x))\partial\phi\right\}
\nonumber\\
   &\qquad
   {}-\ln\left|\prod_{x\in\Lambda}\left\{1-W''(\phi(x))\right\}\mp1\right|.
\label{threextwentyfive}
\end{align}
Note that the fermion determinant~(\ref{threextwentyfour}) is real for a real
superpotential~$W$ but it is not necessarily positive definite. We thus must
include the sign of the determinant,
\begin{equation}
   s[\phi]\equiv
   \sign\left(\det\left\{-\partial-W''(\phi)\right\}\right)
   =\sign\left(\prod_{x\in\Lambda}\left\{1-W''(\phi(x))\right\}\mp1\right),
\label{threextwentysix}
\end{equation}
as a reweighting factor in the functional integral. For example, the
expectation value of the Hamiltonian~$H$ is given by the ratio of
\begin{equation}
   \mathcal{N}\int\prod_{x\in\Lambda}d\phi(x)\,Hs[\phi]\,
   e^{-S_{\text{eff}}[\phi]}
\label{threextwentyseven}
\end{equation}
to the partition function
\begin{equation}
   \mathcal{Z}=\mathcal{N}\int\prod_{x\in\Lambda}d\phi(x)\,s[\phi]\,
   e^{-S_{\text{eff}}[\phi]}.
\label{threextwentyeight}
\end{equation}
Recall that, when supersymmetry is spontaneously broken, the normalized
expectation value with the periodic boundary condition cannot be defined
because $\mathcal{Z}_{\text{PBC}}=0$. With the antiperiodic boundary
condition, expectation values are always meaningful and the following
substitutions can be made:
\begin{equation}
   \left\langle F(x)^2\right\rangle_{\text{aPBC}}=-1,\qquad
   \langle F(x)\rangle_{\text{aPBC}}=0
\end{equation}
for the auxiliary variable, and
\begin{equation}
   \left\langle
   \overline\psi(x)\left(\partial+W''(\phi(x))\right)\psi(x)
   \right\rangle_{\text{aPBC}}=-1
\end{equation}
and
\begin{equation}
   \left\langle\overline\psi(x)W''(\phi(x))\psi(x)\right\rangle_{\text{aPBC}}
   =\left\langle
   W''(\phi(x))\frac{\prod_{y\neq x\in\Lambda}\left\{1-W''(\phi(y))\right\}}
   {\prod_{z\in\Lambda}\left\{1-W''(\phi(z))\right\}+1}\right\rangle_{\text{aPBC}}
\end{equation}
for fermionic variables. It is then straightforward to implement the hybrid
Monte Carlo algorithm\cite{Duane:1987de} with the effective
action~(\ref{threextwentyfive}) and compute $\langle H(x)\rangle_{\text{aPBC}}$. 

Now, as a definite example in which supersymmetry is spontaneously broken, we
consider\footnote{The potential energy $V(\phi)=W'(\phi)^2/2$ is a double-well
type with two minima $V=0$ at $\phi=0$ and $\phi=-m/g$, and the height of the
potential barrier is $m^4/(32g^2)=m/(32\lambda^2)$. From this, the spontaneous
supersymmetry breaking in the present system will be rather difficult to be
observed numerically for \emph{weak\/} couplings for which the supersymmetry
breaking is caused mainly by quantum tunneling.}
\begin{equation}
   W_{\text{continuum}}=\frac{1}{2}m\phi_{\text{continuum}}^2
   +\frac{1}{3}g\phi_{\text{continuum}}^3.
\label{threexthirtytwo}
\end{equation}
Since the parameter~$m$ has the mass dimension~$1$, we will measure all
dimensionful quantities in units of~$m$. For example, the lattice spacing is
measured by the dimensionless combination~$am$. If we introduce the
dimensionless coupling constant as
\begin{equation}
   \lambda\equiv\frac{g}{m^{3/2}},
\end{equation}
the superpotential in terms of the lattice variables reads
\begin{equation}
   W(\phi(x))=\frac{1}{2}(am)\phi(x)^2+\frac{1}{3}(am)^{3/2}\lambda\phi(x)^3.
\label{threexthirtyfour}
\end{equation}

Before proceeding to the numerical study, it is instructive to see how our
method works for the free theory $\lambda=0$, for which supersymmetry is
\emph{not\/} spontaneously broken in the continuum theory and the lattice
model~(\ref{threexnineteen}) is solvable. From Eqs.~(\ref{threexthirtyfour})
and~(\ref{threextwentysix}), we see that the partition function
$\mathcal{Z}$~(\ref{threextwentyeight}) is nonzero for both boundary
conditions (for $a\neq0$). It is therefore meaningful to consider the
expectation value for both boundary conditions and it is not difficult to see
that
\begin{align}
  &\langle H(x)\rangle_{\text{PBC}}=0,
\\
  &\langle H(x)\rangle_{\text{aPBC}}
   =\frac{m(1-am)^{\beta/a-1}}
   {1-(1-am)^{\beta/a}}+\frac{m(1-am)^{\beta/a-1}}{1+(1-am)^{\beta/a}}.
\label{threexthirtysix}
\end{align}
In the continuum limit $a\to0$, the latter, in fact, reproduces the expectation
value of the energy in this supersymmetric harmonic oscillator:
\begin{equation}
   \lim_{a\to0}\langle H(x)\rangle_{\text{aPBC}}
   =\frac{me^{-\beta m}}{1-e^{-\beta m}}+\frac{me^{-\beta m}}{1+e^{-\beta m}}.
\label{threexthirtyseven}
\end{equation}
We thus have, in the large-time limit $\beta\to\infty$,
\begin{equation}
   E_0=\lim_{\beta\to\infty}
   \lim_{a\to0}\langle H(x)\rangle_{\text{aPBC}}=0,
\end{equation}
and infer that supersymmetry is not spontaneously broken.

Now we turn to the Monte Carlo study of the model~(\ref{threexthirtyfour}) with
$\lambda\neq0$ (supersymmetry is dynamically broken in the target continuum
theory). In the following results, we set $\lambda=10$ and, for each set of
parameters, we used $10^4$ statistically independent configurations.

\begin{figure}
\centerline{\includegraphics[width=0.8\textwidth]{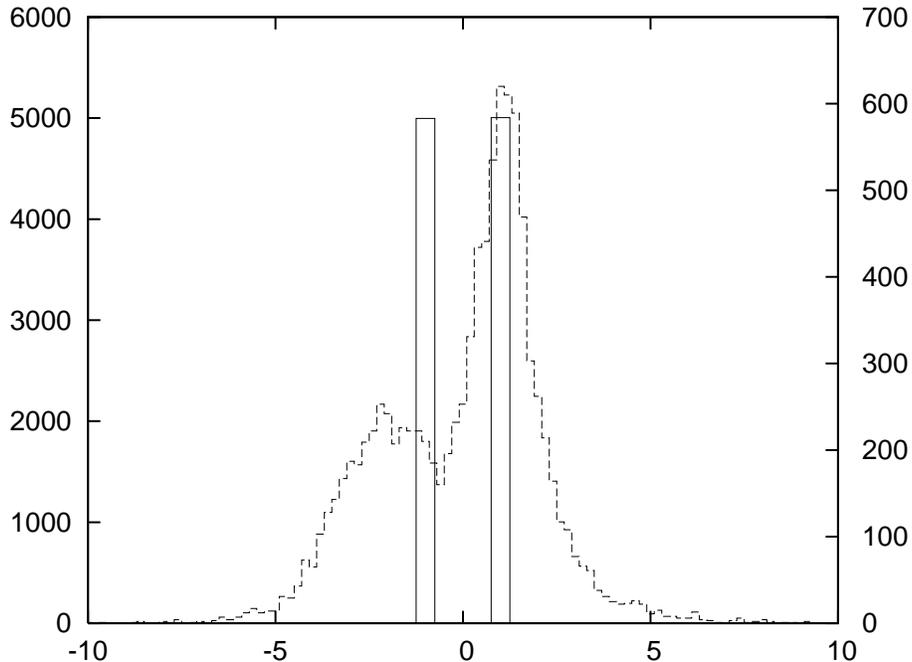}}
\caption{Histogram of the sign of the fermion determinant~$s[\phi]$
(\ref{threextwentysix}) (bold line; left axis) and the Hamiltonian reweighted
by the sign of the determinant $Hs[\phi]$ (broken line; right axis) for the
model~(\ref{threexthirtyfour}) with the periodic boundary condition.
$\lambda=10$. The lattice spacing is $am=0.1$ and the physical temporal size of
the system is $\beta m=1.6$. The number of configurations is $10^4$.}
\label{fig:1}
\end{figure}
First, let us see the case of the periodic boundary condition. For this
boundary condition, the sign of the fermion determinant~(\ref{threextwentysix})
may change depending on the configuration, and in fact, as Fig.~\ref{fig:1}
shows, positive and negative fermion determinants appear at almost equal rates.
This implies that the partition function~(\ref{threextwentyeight}), that is,
the Witten index~(\ref{twoxtwo}), is almost zero. This is perfectly in
accord with the fact that supersymmetry is spontaneously broken in the target
theory. In Fig.~\ref{fig:1}, we plotted also the distribution of the
Hamiltonian~(\ref{threextwentytwo}) reweighted by the sign of the determinant,
$Hs[\phi]$. It spreads on negative as well as positive sides and the average
($\simeq-0.003$) is consistent with zero within statistical error
($\simeq0.02$). This is again consistent with the fact that the average of the
Hamiltonian~(\ref{threextwentyseven}) is merely the $\beta$-derivative of the
Witten index, Eq.~(\ref{twoxthree}). This assures us of the validity of our
method because the construction of $H$ ensures Eq.~(\ref{twoxthree}), as shown
in Eq.~(\ref{threextwentythree}).

\begin{figure} 
\centerline{\includegraphics[width=0.8\textwidth]{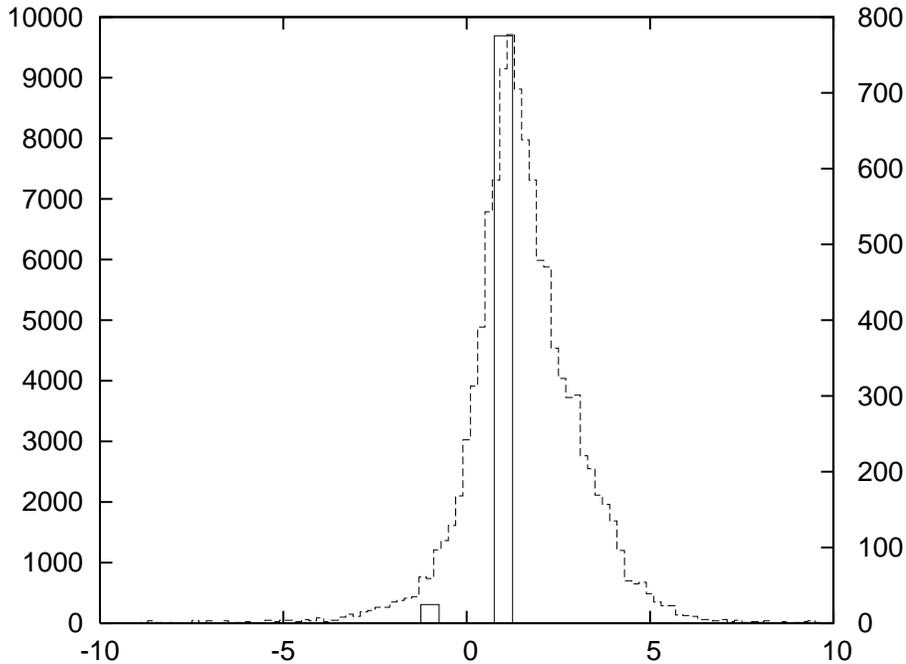}}
\caption{Histogram of the sign of the fermion determinant~$s[\phi]$
(\ref{threextwentysix}) (bold line; left axis) and the Hamiltonian reweighted
by the sign of the determinant $Hs[\phi]$ (broken line; right axis) for the
model~(\ref{threexthirtyfour}) with the antiperiodic boundary condition.
$\lambda=10$. The lattice spacing is $am=0.1$ and the physical temporal size of
the system is $\beta m=1.6$. The number of configurations is $10^4$.}
\label{fig:2}
\end{figure}

If we switch the boundary condition to antiperiodic, things drastically
change. As Fig.~\ref{fig:2} shows, now the distribution of the sign is
significantly asymmetric and the partition function~(\ref{threextwentyeight})
becomes nonvanishing. This implies that we can give a definite meaning for the
expectation value normalized by the partition
function~$\mathcal{Z}_{\text{aPBC}}$.
In this way, we see numerically that the effect of the boundary condition
indeed survives even for large temporal size ($\beta m=1.6$ is actually a large
size with the present value of the coupling constant; see below) when
supersymmetry is spontaneously broken.

\begin{figure} 
\centerline{\includegraphics[width=0.8\textwidth]{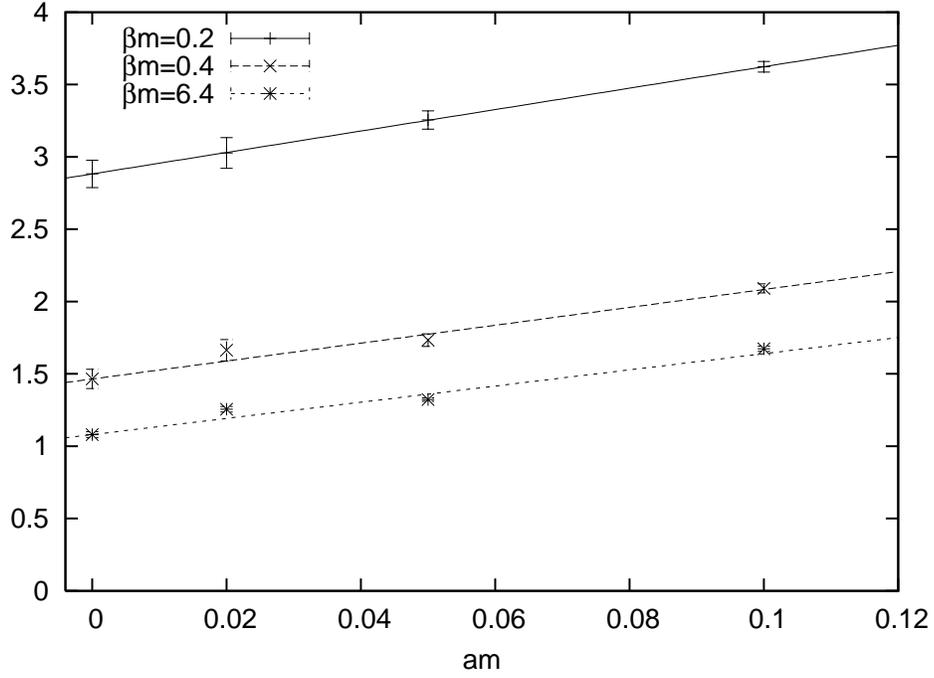}}
\caption{Linear extrapolations of $\langle H(x)\rangle_{\text{aPBC}}/m$ to the
continuum $a=0$ for various values of $\beta m$. $\lambda=10$. The errors are
only statistical ones.}
\label{fig:3}
\end{figure}

\begin{figure} 
\centerline{\includegraphics[width=0.8\textwidth]{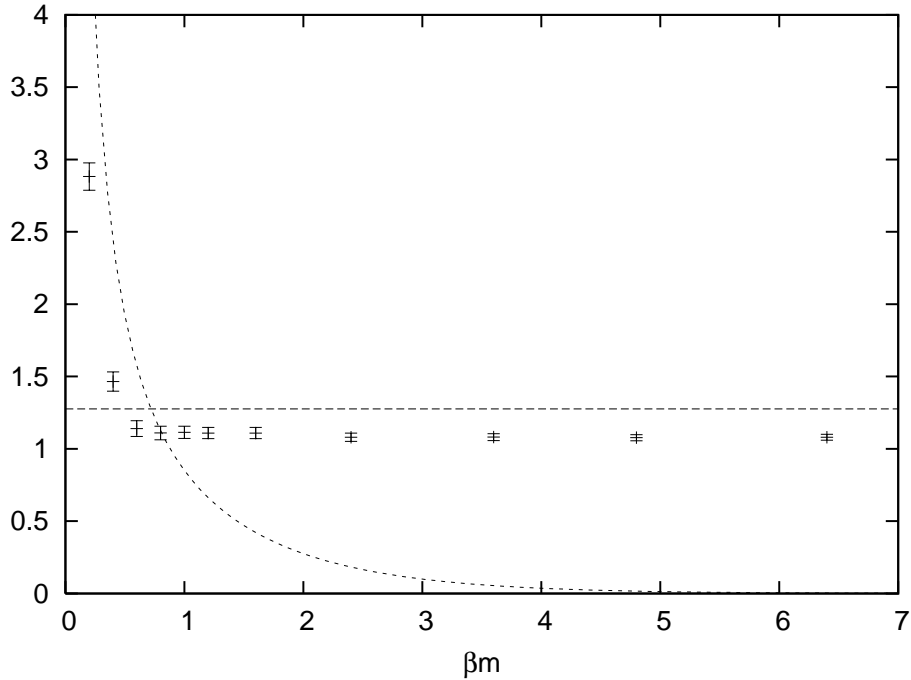}}
\caption{The continuum limit of the expectation value of the Hamiltonian,
$\lim_{a\to0}\langle H(x)\rangle_{\text{aPBC}}/m$, as a function of the physical
temporal size of the system~$\beta m$. $\lambda=10$. The errors are only
statistical ones. We have also plotted the exact ground-state energy
$E_0/m=1.27616$ and the analytic expression for the $\lambda=0$ case,
Eq.~(\ref{threexthirtyseven}).}
\label{fig:4}
\end{figure}

\begin{figure} 
\centerline{\includegraphics[width=0.8\textwidth]{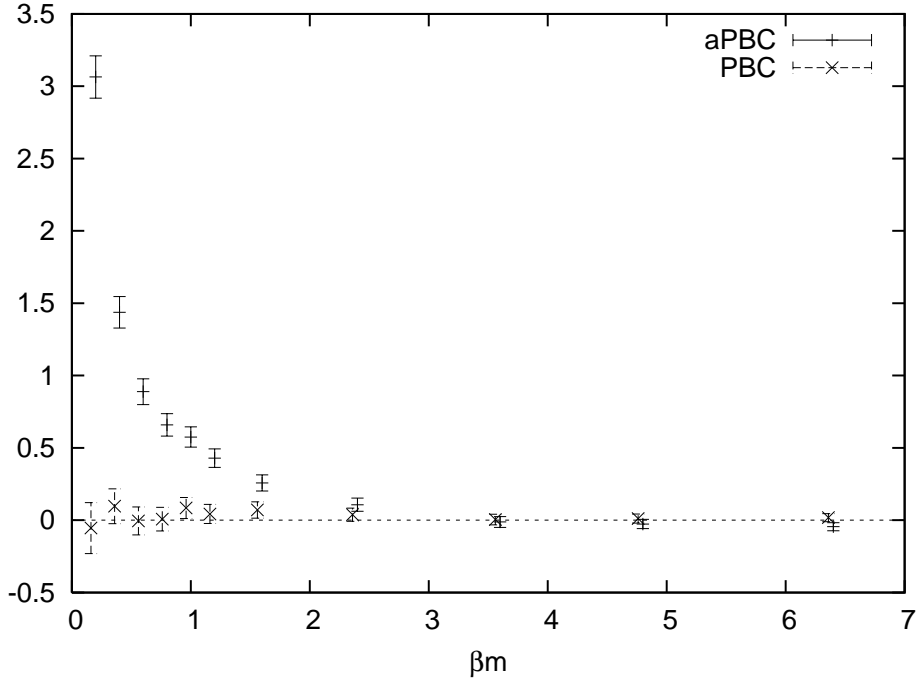}}
\caption{Continuum limit of the expectation values of the Hamiltonian,
$\lim_{a\to0}\langle H(x)\rangle_{\text{aPBC}}/m$ and
$\lim_{a\to0}\langle H(x)\rangle_{\text{PBC}}/m$, as a function of the physical
temporal size of the system~$\beta m$. The errors are only statistical ones.}
\label{fig:5}
\end{figure}

With the antiperiodic boundary condition, we then measure the expectation
value of the Hamiltonian as a function of the temporal size of the
system~$\beta$. For various values of~$\beta m$, we measured
$\langle H(x)\rangle_{\text{aPBC}}/m$ for lattice spacings $am=0.1$, $0.05$
and~$0.02$. The number of configurations is $10^4$ for each set of parameters.
Then, as shown in Fig.~\ref{fig:3}, we extrapolate
$\langle H(x)\rangle_{\text{aPBC}}/m$ to the continuum $a=0$ by a linear
$\chi^2$-fit.\footnote{%
The statistical errors in Fig.~\ref{fig:3} are one standard deviation. The
errors in the linear $\chi^2$-extrapolation were estimated from the range of
fitting parameters that corresponds to a unit variation of $\chi^2$. These
remarks also apply to the results in Fig.~\ref{fig:5}.}%
\footnote{We found that a quadratic function of the form
$\alpha(am)^2+\beta$, which gives a somewhat larger
$\langle H(x)\rangle_{\text{aPBC}}/m$ at $a=0$, provides a better fit.
Although this form of the fit function might be suggested theoretically (i.e.,
the residual lattice artifact is $O(a^2)$ instead of $O(a)$), we stick to a
simple linear fit to avoid a possible criticism that nonzero $E_0$ is an
artificial consequence of the fit.}
In Fig.~\ref{fig:4}, we plot the continuum limit of the expectation
value $\lim_{a\to0}\langle H(x)\rangle_{\text{aPBC}}/m$ as a function of the
physical temporal size of the system~$\beta m$.
For $\beta m\gtrsim1$,\footnote{For strong couplings $\lambda\gg1$, it is easy
to see that energy eigenvalues of the present system scale as
$\lambda^{2/3}m$. Thus the difference between the first excited state and the
ground-state would be $\lambda^{2/3}m$ times a number of $O(1)$. This
observation suggests that the expectation value of the Hamiltonian (with the
antiperiodic boundary condition) exponentially approaches the asymptotic value
at $\beta m=\infty$ around $\beta m\gtrsim\lambda^{-2/3}\simeq0.2$ for
$\lambda=10$.} we have $\lim_{a\to0}\langle H(x)\rangle_{\text{aPBC}}\simeq1.1m$
and, from this, we infer that supersymmetry is spontaneously
broken.\footnote{In Fig.~\ref{fig:4}, we plotted also the exact ground-state
energy $E_0/m=1.27616$ that was obtained by numerically diagonalizing the
corresponding Hamiltonian operator (we used the method of
Ref.~\citen{Balsa:1984eg}). The discrepancy of our Monte Carlo results for
$\beta m\gtrsim1$ with this exact result can be understood as a systematic
error associated with a linear extrapolation to the continuum limit.}
Indeed, this is the correct answer.

Next, as an interacting case in which supersymmetry is \emph{not\/}
spontaneously broken, we consider
\begin{equation}
   W_{\text{continuum}}=\frac{1}{4}m^2\phi_{\text{continuum}}^4.
\end{equation}
In this case, we observed that $s[\phi]$~(\ref{threextwentysix}) has almost
always a definite sign for both boundary conditions and thus both
$\langle H(x)\rangle_{\text{PBC}}$ and $\langle H(x)\rangle_{\text{aPBC}}$ can be
considered. In Fig.~\ref{fig:5}, we plot the continuum limit of these
quantities as a function of $\beta m$. The results are obtained by
extrapolation to the continuum $a=0$ with a linear $\chi^2$-fit (like
Fig.~\ref{fig:3}) of data computed at $am=0.1$ and $0.05$. The number of
configurations is $10^4$ for each set of parameters.
The figure shows that, in this case for which supersymmetry is not
spontaneously broken, $\lim_{a\to0}\langle H(x)\rangle_{\text{PBC}}$ is
consistent with zero for all temporal sizes (recall Eq.~(\ref{twoxfour});
in the present model, $\Tr(-1)^F=1$) and
$\lim_{a\to0}\langle H(x)\rangle_{\text{aPBC}}$ approaches zero as the temporal
size of the system is increased. From this, we conclude that $E_0=0$ is within
the error.

In summary, we have observed that our method works perfectly well in the
present supersymmetric quantum mechanics. One can certainly observe whether
or not the dynamical supersymmetry breaking takes place by our method.

\section{Two-dimensional $\mathcal{N}=(2,2)$ super Yang-Mills theory}

The two-dimensional $\mathcal{N}=(2,2)$ super Yang-Mills theory is obtained by
a dimensional reduction of the four-dimensional $\mathcal{N}=1$ super
Yang-Mills theory.\footnote{In what follows, we assume that the gauge group is
$SU(N_c)$.} This seemingly simple supersymmetric system, however, defies a
straightforward low-energy description for several reasons. First, two global
$U(1)$ symmetries in this system cannot be spontaneously broken in two
dimensions and a description by using the Nambu-Goldstone fields is
impossible.\footnote{Nevertheless, it is possible to show that a correlation
function of Noether currents associated with the $U(1)$ symmetries possesses
a massless pole, to all orders of perturbation theory.\cite{Fukaya:2006mg}}
Second, there is no controllable parameter, other than the number of colors
$N_c$ of the gauge group~$SU(N_c)$. (The two-dimensional gauge coupling~$g$
simply provides a mass scale, just like~$\Lambda_{\text{QCD}}$.) The $1/N_c$
expansion is nontrivial because the gaugino and scalars belong to the adjoint
representation. Finally, the classical potential energy of scalar fields
possesses \emph{noncompact\/} flat directions and there are an infinite number
of degenerated classical vacua. This classical degeneracy is not lifted upon
quantum corrections to all orders of perturbation theory.

In our present context, the last point above (noncompact flat directions in
the classical potential) is an obstruction to the determination of the Witten
index. In the weak coupling approximation, zero-momentum modes without
potential (constant degrees of freedom along flat directions) produce a
continuous spectrum starting at zero. This makes the counting of zero-energy
states, and thus the determination of the Witten index in the weak coupling
approximation, awkward. A similar situation arises in the three-dimensional
$\mathcal{N}=2$ super Yang-Mills theory (that can also be obtained by a
dimensional reduction of the four-dimensional $\mathcal{N}=1$ super Yang-Mills
theory). However, in this three-dimensional model, if the gauge group is
$SU(N_c)$, one eliminate zero-momentum bosonic modes by imposing the twisted
boundary conditions for two spatial directions and obtain
$\Tr(-1)^F=1$.\cite{Affleck:1982as}\footnote{Incidentally, this is a good
example of the Witten index generally not being preserved under dimensional
reduction, because $\Tr(-1)^F=N_c$ for the four-dimensional $\mathcal{N}=1$
super Yang-Mills theory. See also \S~4 of Ref.~\citen{de Wit:1988ct}.}
This trick of the twisted boundary conditions, unfortunately, does not work in
two dimensions. The correct value of the Witten index, or even whether it is
well defined or not, is therefore unknown for the two-dimensional
$\mathcal{N}=(2,2)$ super Yang-Mills theory. It is consequently not known
whether supersymmetry is dynamically broken in this system.

Under this situation, Hori and Tong\cite{Hori:2006dk} conjectured that
dynamical supersymmetry breaking occurs in this system, on the basis of the
counting of the number of ground-states in the two-dimensional
$\mathcal{N}=(2,2)$ supersymmetric gauge theory with fundamental chiral
multiplets, combined with a decoupling argument. In what follows, we
numerically investigate this possibility of dynamical supersymmetry breaking
by directly measuring the ground-state energy density in Euclidean lattice
gauge theory.

\subsection{Hamiltonian density in the continuum theory}
The Lagrangian density of the two-dimensional $\mathcal{N}=(2,2)$ super
Yang-Mills theory in the Minkowski spacetime, in terms of the twisted basis of
spinors,\cite{Witten:1988ze,Witten:1990bs} is given by
\begin{align}
   \mathcal{L}&=\frac{1}{g^2}
   \tr\biggl\{
   -\frac{1}{4}[\phi,\overline\phi]^2-H^2+2HF_{01}
   +D_0\phi D_0\overline\phi-D_1\phi D_1\overline\phi
\nonumber\\
   &\qquad\qquad{}
   +\frac{1}{4}\eta[\phi,\eta]+\chi[\phi,\chi]
   -\psi_\mu[\overline\phi,\psi_\mu]
   +2i\chi(iD_0\psi_1+D_1\psi_0)
   -\psi_0D_0\eta-i\psi_1D_1\eta
   \biggr\},
\label{fourxone}
\end{align}
where all fields are $SU(N_c)$ Lie algebra valued and scalar fields $\phi$
and~$\overline\phi$ are combinations of two real scalar fields, $\phi=X_2+iX_3$
and $\overline\phi=X_2-iX_3$, respectively.
$F_{01}=\partial_0A_1-\partial_1A_0+i[A_0,A_1]$ is the field strength in two
dimensions. The covariant derivatives $D_\mu$ are defined with respect to the
adjoint representation $D_\mu\varphi=\partial_\mu\varphi+i[A_\mu,\varphi]$ for
any field~$\varphi$. The index~$\mu$ runs over 0 and~1. Note that, in the above
convention, the bosonic fields $A_\mu$, $\phi$ and~$\overline\phi$ have the
mass dimension~1 and the fermionic fields $\psi_\mu$, $\chi$ and~$\eta$ have
the mass dimension~$3/2$, because the gauge coupling constant in two
dimensions~$g$ has the mass dimension~1.

The action $\int d^2x\,\mathcal{L}$ with the periodic boundary condition is
invariant under four supersymmetry transformations. Among them, what is
relevant to us is $Q$ and $Q_0$. The $Q$-transformation is given by
\begin{align}
   &QA_0=i\psi_0,&& Q\psi_0=D_0\phi,
\nonumber\\
   &QA_1=\psi_1,&& Q\psi_1=iD_1\phi,
\nonumber\\
   &Q\phi=0,&&
\nonumber\\
   &Q\chi=H,&& QH=[\phi,\chi],
\nonumber\\
   &Q\overline\phi=\eta,&& Q\eta=[\phi,\overline\phi]
\end{align}
and $Q_0$ is (see, for example, Ref.~\citen{Kato:2003ss})
\begin{align}
   &Q_0A_0=\frac{i}{2}\eta,&& Q_0\eta=-2D_0\overline\phi,
\nonumber\\
   &Q_0A_1=-\chi,&& Q_0\chi=iD_1\overline\phi,
\nonumber\\
   &Q_0\overline\phi=0,&&
\nonumber\\
   &Q_0\psi_1=H-2F_{01},&&
   Q_0H=-[\overline\phi,\psi_1]-2D_0\chi-iD_1\eta.   
\nonumber\\
   &Q_0\phi=-2\psi_0,&& Q_0\psi_0=\frac{1}{2}[\overline\phi,\phi].
\end{align}
One then finds that these transformations satisfy
\begin{equation}
   Q^2=\delta_\phi,\qquad Q_0^2=-\delta_{\overline\phi},\qquad
   \{Q,Q_0\}=-2\partial_0-2i\delta_{A_0},
\label{fourxfour}
\end{equation}
where $\delta_\varphi$ denotes the infinitesimal gauge transformation with the
parameter~$\varphi$. These differ from the off-shell supersymmetry algebra in
the twisted basis, $Q^2=Q_0^2=0$ and $\{Q,Q_0\}=-2\partial_0$, by gauge
transformations because we are working with the Wess-Zumino gauge.

A crucial property of this system, which allows a simple lattice formulation,
is that the action~$S=\int d^2x\,\mathcal{L}$ is
$Q$-exact.\cite{Witten:1988ze,Witten:1990bs}
\begin{equation}
   S=Q\frac{1}{g^2}
   \int d^2x\,\tr\left\{
   -\frac{1}{4}\eta[\phi,\overline\phi]+2\chi F_{01}-\chi H
   +\psi_0D_0\overline\phi+i\psi_1D_1\overline\phi
   \right\}
\end{equation}
In this form, with the relations~(\ref{fourxfour}), the invariance of the
action under $Q$ and $Q_0$ transformations is easily seen.\footnote{Note that
$S$ can further be written as
$S=QQ_0\frac{1}{g^2}\int d^2x\,\tr\left\{
-\frac{1}{2}\phi D_0\overline\phi-\psi_1\chi\right\}$.}

Now, from the Lagrangian density~(\ref{fourxone}), we obtain the Hamiltonian
density~$\mathcal{H}$ by the Legendre transformation. After (trivially)
eliminating redundant fields by using second-class constraints, $\mathcal{H}$
is given, in terms of fields in the Lagrangian, by
\begin{align}
   \mathcal{H}&=\frac{1}{g^2}
   \tr\biggl\{
   \frac{1}{4}[\phi,\overline\phi]^2+H^2
   +D_0\phi D_0\overline\phi+D_1\phi D_1\overline\phi
\nonumber\\
   &\qquad\qquad{}
   -\frac{1}{4}\eta[\phi,\eta]-\chi[\phi,\chi]
   +\psi_\mu[\overline\phi,\psi_\mu]
   -2i\chi D_1\psi_0+i\psi_1D_1\eta\biggr\}
\nonumber\\
   &\qquad{}
   -2\tr\left\{A_0\mathcal{G}\right\},
\label{fourxsix}
\end{align}
up to a spatial total derivative, where $\mathcal{G}$ is the Gauss-law
constraint:
\begin{equation}
   \mathcal{G}=\frac{1}{g^2}
   \left\{D_1H+\frac{i}{2}[\phi,D_0\overline\phi]
   +\frac{i}{2}[\overline\phi,D_0\phi]
   +i\left\{\psi_1,\chi\right\}
   +\frac{i}{2}\left\{\eta,\psi_0\right\}\right\}.
\end{equation}
From the off-shell supersymmetry algebra $\{Q,Q_0\}=-2\partial_0$, one might
expect that the relation $Q\mathcal{J}_0^0/2=\mathcal{H}$ holds, where
$\mathcal{J}_0^0$ is the time component of the Noether current associated with
the $Q_0$-symmetry
\begin{equation}
   \mathcal{J}_0^0=\frac{1}{g^2}
   \tr\left\{\frac{1}{2}\eta[\phi,\overline\phi]
   +2\chi H+2\psi_0D_0\overline\phi-2i\psi_1D_1\overline\phi\right\}.
\end{equation}
In reality,
\begin{equation}
   \frac{1}{2}Q\mathcal{J}_0^0
   =\mathcal{H}
   +2\tr\left\{A_0\mathcal{G}\right\}
   +\frac{1}{g^2}
   \tr\left\{\psi_0\left(-2[\overline\phi,\psi_0]+2iD_1\chi
   -D_0\eta\right)\right\}
\label{fourxnine}
\end{equation}
up to a spatial total derivative. Compare this with Eq.~(\ref{twoxeight}). In
Eq.~(\ref{fourxnine}), the last two terms are proportional to classical
equations of motion and they can be expressed as
$A_0^a\frac{\delta}{\delta A_0^a}S$ and
$\psi_0^a\frac{\delta}{\delta\psi_0^a}S$.\footnote{We define color components
of fields as $\varphi=\sum_{a=1}^{N_c^2-1}\varphi^aT^a$, where $T^a$ are
generators of $SU(N_c)$.}
The expectation value of these two expressions may be obtained (after gauge
fixing) as a Jacobian associated with the transformations
$A_0\to A_0+\alpha A_0$ and~$\psi_0\to\psi_0+\alpha\psi_0$, respectively. Such a
Jacobian is generally ultraviolet divergent.\footnote{With a lattice
regularization, for example, the one described in the next subsection, the
expectation value of the latter is~$-i(N_c^2-1)/a^2$. The expectation value of
the former is $+i(N_c^2-1)/a^2$ after gauge fixing and thus, quite
interestingly, the expectation value of the last two terms of
Eq.~(\ref{fourxnine}) vanishes.}

Thus, in view of Eq.~(\ref{fourxnine}) and following our general prescription,
we adopt $\mathcal{H}\equiv Q\mathcal{J}_0^0/2$ as the Hamiltonian density in
our functional integral formulation. As can be seen from Eqs.~(\ref{fourxnine})
and~(\ref{fourxsix}), this $\mathcal{H}$ is, moreover, gauge invariant.

Thus, going to the Euclidean space by $x_0\to-ix_0$, $A_0\to iA_0$,
$D_0\to iD_0$ and $\mathcal{L}\to-\mathcal{L}$, we have the Euclidean action
\begin{equation}
   S=Q\frac{1}{g^2}
   \int d^2x\,\tr\left\{
   \frac{1}{4}\eta[\phi,\overline\phi]-i\chi\Phi+\chi H
   -i\psi_\mu D_\mu\overline\phi\right\},
\label{fourxten}
\end{equation}
where $\Phi\equiv 2F_{01}$ and
\begin{align}
   &QA_\mu=\psi_\mu,&&Q\psi_\mu=iD_\mu\phi,
\nonumber\\
   &Q\phi=0,&&
\nonumber\\
   &Q\chi=H,&&QH=[\phi,\chi],
\nonumber\\
   &Q\overline\phi=\eta,&&Q\eta=[\phi,\overline\phi],
\label{fourxeleven}
\end{align}
and the Hamiltonian density
\begin{equation}
   \mathcal{H}\equiv Q\frac{1}{g^2}\tr\left\{
   \frac{1}{4}\eta[\phi,\overline\phi]
   +\chi H
   +i\psi_0D_0\overline\phi
   -i\psi_1D_1\overline\phi\right\}.
\label{fourxtwelve}
\end{equation}
These are the basic relations for our Euclidean lattice formulation.

\subsection{Manifestly $Q$-invariant lattice formulation}
This subsection is a brief summary of a lattice formulation of the
two-dimensional $\mathcal{N}=(2,2)$ super Yang-Mills theory proposed in
Ref.~\citen{Sugino:2004qd}. For full details, we refer the reader to
Ref.~\citen{Sugino:2004qd}. We consider a two-dimensional rectangular lattice
of the physical size $\beta\times L$:
\begin{equation}
   \Lambda=\left\{x\in a\mathbb{Z}^2\mid 0\leq x_0<\beta,\,\,
   0\leq x_1<L\right\},
\end{equation}
where $a$~denotes the lattice spacing. All fields except the gauge potentials
are put on sites and the gauge field is expressed by the compact link
variables~$U(x,\mu)\in SU(N_c)$.

As a lattice transcription of the $Q$-transformation~(\ref{fourxeleven}), we
define
\begin{align}
   &QU(x,\mu)=i\psi_\mu(x)U(x,\mu),
\nonumber\\
   &Q\psi_\mu(x)=i\psi_\mu(x)\psi_\mu(x)
   -i\left(\phi(x)-U(x,\mu)\phi(x+a\hat\mu)U(x,\mu)^{-1}\right),
\nonumber\\
   &Q\phi(x)=0,
\nonumber\\
   &Q\chi(x)=H(x),\qquad QH(x)=[\phi(x),\chi(x)],
\nonumber\\
   &Q\overline\phi(x)=\eta(x),\qquad Q\eta(x)=[\phi(x),\overline\phi(x)]
\label{fourxforteen}
\end{align}
($\hat\mu$ implies a unit vector in the $\mu$-direction).
It can be confirmed that $Q^2=\delta_\phi$, where $\delta_\phi$ is an
infinitesimal gauge transformation \emph{on the lattice\/} with the
parameter~$\phi(x)$. The lattice action is then defined by an expression
analogous to Eq.~(\ref{fourxten}):
\begin{equation}
   S=Qa^2\sum_{x\in\Lambda}\left(
   \mathcal{O}_1(x)+\mathcal{O}_2(x)+\mathcal{O}_3(x)
   +\frac{1}{a^4g^2}\tr\left\{\chi(x)H(x)\right\}\right),
\label{fourxfifteen}
\end{equation}
where
\begin{align}
   &\mathcal{O}_1(x)
   =\frac{1}{a^4g^2}
   \tr\left\{\frac{1}{4}\eta(x)[\phi(x),\overline\phi(x)]\right\},
\label{fourxsixteen}\\
   &\mathcal{O}_2(x)
   =\frac{1}{a^4g^2}
   \tr\left\{-i\chi(x)\hat\Phi(x)\right\},
\label{fourxseventeen}\\
   &\mathcal{O}_3(x)
   =\frac{1}{a^4g^2}
   \tr\left\{i\sum_{\mu=0}^1\psi_\mu(x)
   \left(\overline\phi(x)
   -U(x,\mu)\overline\phi(x+a\hat\mu)U(x,\mu)^{-1}\right)\right\}.
\label{fourxeighteen}
\end{align}
In Eq.~(\ref{fourxseventeen}), $\hat\Phi(x)$ is a lattice counterpart of the
field strength and is defined from the plaquette variables
\begin{equation}
    U(x,0,1)=U(x,0)U(x+a\hat0,1)U(x+a\hat1,0)^{-1}U(x,1)^{-1}
\label{fourxnineteen}
\end{equation}
as
\begin{equation}
   \hat\Phi(x)
   =\frac{\Phi(x)}{1-\frac{1}{\epsilon^2}\left\|1-U(x,0,1)\right\|^2},
   \qquad
   \Phi(x)=-i\left[U(x,0,1)-U(x,0,1)^{-1}\right],
\label{fourxtwenty}
\end{equation}
where the matrix norm is
\begin{equation}
   \|A\|=\left[\tr\left\{AA^\dagger\right\}\right]^{1/2}
\end{equation}
and the constant~$\epsilon$ is chosen in the range
\begin{align}
   &0<\epsilon<2\sqrt{2},\qquad\text{for $N_c=2$, 3, 4},
\label{fourxtwentytwo}
\\
   &0<\epsilon<2\sqrt{N_c}\sin\left(\frac{\pi}{N_c}\right),
   \qquad\text{for $N_c\geq5$}.
\end{align}
From the $Q$-exact form~(\ref{fourxfifteen}) and the nilpotency of $Q$,
$Q^2=\delta_\phi$, the lattice action is manifestly invariant under the
$Q$-transformation~(\ref{fourxforteen}).\footnote{Another interesting property
of the present lattice formulation is that one global $U(1)_R$ symmetry is
manifestly preserved.\cite{Sugino:2004qd}}

After the operation of~$Q$, the lattice action becomes
\begin{equation}
   S=a^2\sum_{x\in\Lambda}\left(
   \sum_{i=1}^4\mathcal{L}_{\text{B}i}(x)+\sum_{i=1}^7\mathcal{L}_{\text{F}i}(x)
   +\frac{1}{a^4g^2}\tr\left\{H(x)-\frac{1}{2}i\hat\Phi_{\text{TL}}(x)\right\}^2
   \right),
\label{fourxtwentyfour}
\end{equation}
where we have noted that only the \emph{traceless part\/} of $\hat\Phi(x)$,
\begin{equation}
   \hat\Phi_{\text{TL}}(x)=\hat\Phi(x)
   -\frac{1}{N_c}\tr\left\{\hat\Phi(x)\right\}\mathbbm{1},
\end{equation}
appears in the action, because the auxiliary field~$H(x)$ is traceless. Each
term of the action density is given by
\begin{align}
   &\mathcal{L}_{\text{B}1}(x)=\frac{1}{a^4g^2}
   \tr\left\{\frac{1}{4}[\phi(x),\overline\phi(x)]^2\right\},
\\
   &\mathcal{L}_{\text{B}2}(x)=\frac{1}{a^4g^2}
   \tr\left\{\frac{1}{4}\hat\Phi_{\text{TL}}(x)^2\right\},
\\
   &\mathcal{L}_{\text{B}3}(x)=\frac{1}{a^4g^2}
   \tr\Biggl\{
   \left(\phi(x)-U(x,0)\phi(x+a\hat0)U(x,0)^{-1}\right)
\nonumber\\
   &\qquad\qquad\qquad\qquad\qquad{}
   \times
   \left(\overline\phi(x)-U(x,0)\overline\phi(x+a\hat0)U(x,0)^{-1}
   \right)\Biggr\},
\\
   &\mathcal{L}_{\text{B}4}(x)=\frac{1}{a^4g^2}
   \tr\Biggl\{
   \left(\phi(x)-U(x,1)\phi(x+a\hat1)U(x,1)^{-1}\right)
\nonumber\\
   &\qquad\qquad\qquad\qquad\qquad{}
   \times
   \left(\overline\phi(x)-U(x,1)\overline\phi(x+a\hat1)U(x,1)^{-1}
   \right)\Biggr\},
\end{align}
and
\begin{align}
   &\mathcal{L}_{\text{F}1}(x)=\frac{1}{a^4g^2}
   \tr\left\{-\frac{1}{4}\eta(x)[\phi(x),\eta(x)]\right\},
\\
   &\mathcal{L}_{\text{F}2}(x)=\frac{1}{a^4g^2}
   \tr\left\{-\chi(x)[\phi(x),\chi(x)]\right\},
\\
   &\mathcal{L}_{\text{F}3}(x)=\frac{1}{a^4g^2}
   \tr\left\{-\psi_0(x)\psi_0(x)
   \left(\overline\phi(x)+U(x,0)\overline\phi(x+a\hat0)U(x,0)^{-1}
   \right)\right\},
\\
   &\mathcal{L}_{\text{F}4}(x)=\frac{1}{a^4g^2}
   \tr\left\{-\psi_1(x)\psi_1(x)
   \left(\overline\phi(x)+U(x,1)\overline\phi(x+a\hat1)U(x,1)^{-1}
   \right)\right\},
\\
   &\mathcal{L}_{\text{F}5}(x)=\frac{1}{a^4g^2}
   \tr\left\{i\chi(x)Q\hat\Phi(x)\right\},
\\
   &\mathcal{L}_{\text{F}6}(x)=\frac{1}{a^4g^2}
   \tr\left\{-i\psi_0(x)
   \left(\eta(x)-U(x,0)\eta(x+a\hat0)U(x,0)^{-1}\right)\right\}.
\\
   &\mathcal{L}_{\text{F}7}(x)=\frac{1}{a^4g^2}
   \tr\left\{-i\psi_1(x)
   \left(\eta(x)-U(x,1)\eta(x+a\hat1)U(x,1)^{-1}\right)\right\}.
\end{align}
Note that all lattice fields in the above expressions are dimensionless. For
all fields other than fermionic fields, periodic boundary conditions
on~$\Lambda$ are assumed. For fermionic fields,
$\psi\equiv(\psi_\mu,\chi,\eta)$, depending on whether the temporal boundary
condition is periodic (PBC) or antiperiodic (aPBC), we set
\begin{equation}
   \psi(x_0=\beta,x_1)=\begin{cases}
   +\psi(x_0=0,x_1)&\text{for the periodic boundary condition},\\
   -\psi(x_0=0,x_1)&\text{for the antiperiodic boundary condition},
   \end{cases}
\end{equation}
while the spatial boundary condition is always taken to be periodic.

With the lattice action~(\ref{fourxtwentyfour}), the partition function is
defined by
\begin{equation}
   \mathcal{Z}=\mathcal{N}\int d\mu\,e^{-S},
\label{fourxthirtyeight}
\end{equation}
where the integration measure is defined (writing $\phi(x)=X_2(x)+iX_3(x)$
and $\overline\phi(x)=X_2(x)-iX_3(x)$) by
\begin{equation}
   d\mu\equiv\prod_{x\in\Lambda}\left(\prod_{\mu=0}^1d U(x,\mu)\right)
   \prod_{a=1}^{N_c^2-1}d X_2^a(x)\,d X_3^a(x)\,d H^a(x)
   \left(\prod_{\mu=0}^1d\psi_\mu^a(x)\right)
   d\chi^a(x)\,d\eta^a(x)
\end{equation}
in terms of color components of fields. $d U(x,\mu)$ is the standard Haar
measure. Note that the integration over the auxiliary field~$H(x)$ can readily
be performed because it is gaussian. The invariance of this measure under the
$Q$-transformation is noted in Ref.~\citen{Sugino:2006uf}.

The denominator in Eq.~(\ref{fourxtwenty}) needs an explanation. Without that
factor, the lattice action for the gauge field is the ``double-winding
plaquette type''\cite{Elitzur:1982vh} and the action possesses many degenerate
minima which have no continuum counterpart. Because of the denominator of
Eq.~(\ref{fourxtwenty}), the action~(\ref{fourxtwentyfour}) diverges as
$\left\|1-U(x,0,1)\right\|\to\epsilon$ at a certain site~$x$. Precisely
speaking, the above construction of the action is applied only for
configurations with
\begin{equation}
     \left\|1-U(x,0,1)\right\|<\epsilon,\qquad\hbox{for $\forall x\in\Lambda$},
\label{fourxforty}
\end{equation}
and, otherwise, i.e., if there exists $x\in\Lambda$ such that
$\left\|1-U(x,0,1)\right\|\geq\epsilon$, we set
\begin{equation}
   S=+\infty.
\end{equation}
In this way, the domain of functional integral~(\ref{fourxthirtyeight}) is
effectively restricted to configurations specified by Eq.~(\ref{fourxforty}).
It can then be shown that $U(x,\mu)\equiv1$ is (up to gauge transformations) a
unique minimum of the action within the integration domain. This procedure to
solve the problem of degenerate minima moreover does not break the
$Q$-symmetry.\cite{Sugino:2004qd}

With the above construction (and with the periodic boundary condition), one
fermionic symmetry~$Q$ is manifestly preserved on the lattice. The price to pay
is that the Pfaffian of the Dirac operator, resulting from the integration over
fermionic fields, is generally complex. Since the corresponding Pfaffian in the
target continuum theory is real and positive semi-definite, the complex phase
must be a lattice artifact. That is, we expect that the imaginary part of the
lattice Pfaffian diminishes as we approach the continuum limit. We will later
confirm this expectation numerically.

In Refs.~\citen{Sugino:2003yb} and~\citen{Sugino:2004qd}, the restoration of
the invariance under a full set of supersymmetry transformation in the
continuum limit has been argued on the basis of perturbative power counting. It
is certainly desirable to confirm, however, this restoration nonperturbatively
by observing the supersymmetric Ward-Takahashi identities. More definitely, one
should examine the total divergence of two-point (denominator-free) correlation
functions containing the supercurrent; so far this analysis has not yet been
carried out.

\subsection{Lattice transcription of Hamiltonian density}
Now, as the definition of a Hamiltonian density on the lattice, we follow the
prescription $\mathcal{H}(x)\equiv Q\mathcal{J}_0^0(x)/2$ suggested by
Eq.~(\ref{fourxtwelve}), where
\begin{align}
   \mathcal{J}_0^0(x)&\equiv\frac{1}{a^4g^2}
   \tr\biggl\{\frac{1}{2}\eta(x)[\phi(x),\overline\phi(x)]+2\chi(x)H(x)
\nonumber\\
   &\qquad\qquad\qquad{}
   -2i\psi_0(x)
   \left(\overline\phi(x)
   -U(x,0)\overline\phi(x+a\hat0)U(x,0)^{-1}\right)
\nonumber\\
   &\qquad\qquad\qquad{}
   +2i\psi_1(x)
   \left(\overline\phi(x)
   -U(x,1)\overline\phi(x+a\hat1)U(x,1)^{-1}\right)\biggr\}
\end{align}
is a lattice transcription of the Noether current $\mathcal{J}_0^0$ in
Eq.~(\ref{fourxtwelve}). The explicit form of the Hamiltonian density is then
given by
\begin{align}
   \mathcal{H}(x)
   &=\mathcal{L}_{\text{B}1}(x)
   -\mathcal{L}_{\text{B}3}(x)
   +\mathcal{L}_{\text{B}4}(x)
\nonumber\\
   &\qquad{}
   +\mathcal{L}_{\text{F}1}(x)+\mathcal{L}_{\text{F}2}(x)
   -\mathcal{L}_{\text{F}3}(x)+\mathcal{L}_{\text{F}4}(x)
   -\mathcal{L}_{\text{F}6}(x)+\mathcal{L}_{\text{F}7}(x)
\nonumber\\
   &\qquad{}
   +\frac{1}{a^4g^2}\tr\left\{H(x)\right\}^2.
\label{fourxfortythree}
\end{align}
From the $Q$-invariance of the lattice action and of the integration measure,
we thus have
\begin{equation}
   \int_{\text{PBC}}d\mu\,\mathcal{H}(x)\,e^{-S}
   =\int_{\text{PBC}}d\mu\,Q
   \left(\frac{1}{2}\mathcal{J}_0^0(x)\,e^{-S}\right)=0,
\label{fourxfortyfour}
\end{equation}
assuming that the integral $\int_{\text{PBC}}d\mu\,\mathcal{J}_0^0(x)\,e^{-S}$
is finite, and this reproduces the topological property of the Witten index,
Eq.~(\ref{twoxthree}). As already discussed, we regard this property as a
guiding principle for choosing the origin of the energy (density).

We will measure the ground-state (vacuum) energy density~$\mathcal{E}_0$ by
\begin{equation}
   \lim_{\beta\to\infty}\lim_{a\to0}\langle\mathcal{H}(x)\rangle_{\text{aPBC}}
   =\lim_{\beta\to\infty}\lim_{a\to0}
   \frac{\int_{\text{aPBC}}d\mu\,\mathcal{H}(x)\,e^{-S}}
   {\int_{\text{aPBC}}d\mu\,e^{-S}}
   =\mathcal{E}_0
\label{fourxfortyfive}
\end{equation}
and judge that dynamical supersymmetry breaking occurs if $\mathcal{E}_0>0$
and but not if $\mathcal{E}_0=0$.

For Eq.~(\ref{fourxfortythree}), an integration over the auxiliary field~$H(x)$
can be performed to obtain
\begin{align}
   \frac{1}{a^4g^2}\left\langle\tr\left\{H(x)\right\}^2
   \right\rangle_{\text{aPBC}}
   &=\frac{1}{2}(N_c^2-1)\frac{1}{a^2}
   -\frac{1}{a^4g^2}\left\langle
   \tr\left\{\frac{1}{4}\hat\Phi_{\text{TL}}(x)^2\right\}^2
   \right\rangle_{\text{aPBC}}
\nonumber\\
   &=\frac{1}{2}(N_c^2-1)\frac{1}{a^2}
   -\left\langle\mathcal{L}_{\text{B}2}(x)\right\rangle_{\text{aPBC}}.
\end{align}
Thus, in actual numerical simulations, we can (suppressing the subscript
aPBC or PBC) use
\begin{align}
   \langle\mathcal{H}(x)\rangle
   &=\left\langle\mathcal{L}_{\text{B}1}(x)\right\rangle
   -\left\langle\mathcal{L}_{\text{B}2}(x)\right\rangle
   -\left\langle\mathcal{L}_{\text{B}3}(x)\right\rangle
   +\left\langle\mathcal{L}_{\text{B}4}(x)\right\rangle
\nonumber\\
   &\qquad{}
   +\left\langle\mathcal{L}_{\text{F}1}(x)\right\rangle
   +\left\langle\mathcal{L}_{\text{F}2}(x)\right\rangle
   -\left\langle\mathcal{L}_{\text{F}3}(x)\right\rangle
   +\left\langle\mathcal{L}_{\text{F}4}(x)\right\rangle
   -\left\langle\mathcal{L}_{\text{F}6}(x)\right\rangle
   +\left\langle\mathcal{L}_{\text{F}7}(x)\right\rangle
\nonumber\\
   &\qquad{}+\frac{1}{2}(N_c^2-1)\frac{1}{a^2}.
\label{fourxfortyseven}
\end{align}

We can argue that the expectation value (\ref{fourxfortyseven}) with aPBC is
ultraviolet finite and possesses a well defined continuum limit to all orders
of (lattice) perturbation theory. In the present super-renormalizable model,
a possible ultraviolet divergence in Eq.~(\ref{fourxfortyseven}) arises from
either (i)~one-loop two-point functions of bosons contained as sub-diagrams or
(ii)~one-loop diagrams that are formed by a self-contraction of kinetic
terms. The former (i) case is potentially logarithmically divergent but the
divergence is cancelled among boson loops and fermion loops. For PBC, this
cancellation can be shown by using the $Q$-symmetry. For aPBC, the cancellation
still holds because the change in the boundary condition does not influence the
coefficients of the logarithmic divergent pieces. In the latter case (ii), the
divergence is quadratic and it is common to the free theory. We may thus
examine the expectation value in the free theory and find that, even with aPBC,
it is ultraviolet finite for any $\beta>0$. Our numerical study described below
in fact indicates that the continuum limit of the expectation value
$\langle\mathcal{H}(x)\rangle_{\text{aPBC}}$ is well defined.

\subsection{Monte Carlo study}
We numerically studied only the $SU(2)$ gauge group. Our algorithm and
the computation code, which was developed using
FermiQCD/MDP,\cite{DiPierro:2000bd,DiPierro:2005qx} are almost identical to
those in Ref.~\citen{Suzuki:2007jt}. We use the hybrid Monte Carlo
algorithm\cite{Duane:1987de} to generate configurations in the quenched
approximation. The effect of dynamical fermions is later taken into account
by reweighting configurations by the Pfaffian of the Dirac operator. We do not
introduce any mass terms of fermions or bosons that would explicitly break the
$Q$-symmetry. Although this is certainly a brute force method compared with a
standard pseudo-fermion algorithm, its implementation is much simpler and the
validity has been confirmed for one-point Ward-Takahashi
identities.\cite{Suzuki:2007jt}

\begin{figure}
\centerline{\includegraphics[width=0.8\textwidth]{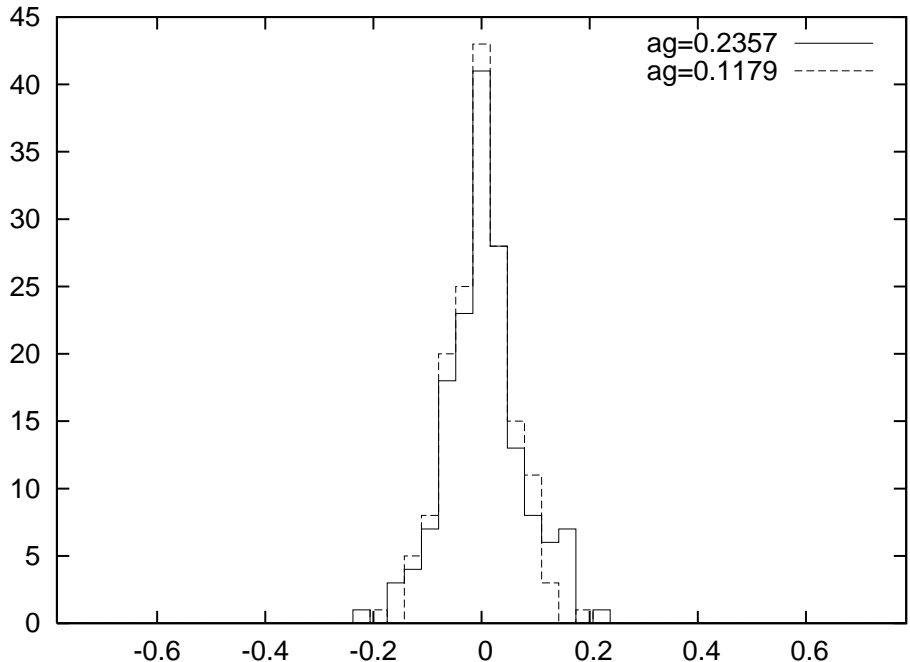}}
\caption{Histogram of the complex phase of the Pfaffian $\Arg(\Pf\{D\})$ in
radians obtained by direct calculation of the Pfaffian for sampled
configurations. The plots are for two entries in the $N_T/N_S=1$ column in
Table~\ref{table:1} and the number of sampled configurations is 160 for both
cases. The boundary condition is antiperiodic.}
\label{fig:6}
\end{figure}

A direct calculation of the Pfaffian is very time-consuming.\footnote{It can be
seen that the algorithm for the Pfaffian (appearing, for example, in
Ref.~\citen{Campos:1999du}) is an $O(n^4)$ process for a $2n\times2n$ matrix,
while the LU decomposition has an $O(n^3)$-process algorithm.} Thus we instead
use a square root of the determinant that is obtained by LU decomposition.
Expressing the determinant of the Dirac operator~$D$ in the form
\begin{equation}
   \det\{D\}=re^{i\theta},\qquad-\pi<\theta\leq\pi,
\end{equation}
we evaluate the Pfaffian by
\begin{equation}
   \Pf\{D\}=\sqrt{r}e^{i\theta/2},
\end{equation}
because $(\Pf\{D\})^2=\det\{D\}$. This prescription, however, reproduces a
correct Pfaffian if and only if $-\pi/2<\Arg(\Pf\{D\})\leq\pi/2$. It is
expected that this inequality is fulfilled in the continuum limit, because the
Pfaffian in the continuum target theory is real and positive semi-definite. A
direct calculation of the Pfaffian over a subset of our configurations
(Fig.~\ref{fig:6}) clearly supports this expectation and justifies the above
prescription. Note that our present lattices are much finer, compared with that
in Ref.~\citen{Suzuki:2007jt} where $ag\geq0.5$.

We stored statistically independent configurations for parameters summarized in
Table~\ref{table:1}, where $N_T$ and $N_S$ are the number of lattice points for
the temporal and spatial directions, respectively. The physical size of the
spatial direction is fixed to be $Lg=\sqrt{2}$.\footnote{Note that
supersymmetry is not broken in the infinite volume if it is not with finite
volume.\cite{Witten:1981nf} \ This fact would justify our study with finite
physical volume.} The parameter $\epsilon$ in Eq.~(\ref{fourxtwentytwo}) is
taken to be $\epsilon = 2.6$. We used the cold start and set all scalar fields
to be zero at the initial configuration. As the initial thermalization, we
discarded the first $10^4$ trajectories and then stored configurations at
every $10^2$ trajectories (the auto-correlation time was 10--20 trajectories).
\begin{table}
\caption{Number of statistically independent configurations we used for
the cases with the antiperiodic boundary condition. $N_T\times N_S=3\times6$
is the minimal-size lattice and $36\times12$ is the maximal. The physical
spatial size is held fixed at $Lg=\sqrt{2}=1.4142$.}
\label{table:1}
\begin{center}
\begin{tabular}{c|c|ccccccc} \hline\hline
\multicolumn{2}{c|}{} & \multicolumn{7}{c}{$N_T/N_S$} \\
\hline
$N_S$ & $ag$ & 0.25 & 0.5 & 1 & 1.5 & 2 & 2.5 & 3 \\
\hline
6     & 0.2357  & ---     & 39,900 & 99,900 & 9,900 &  9,900 & 9,900 & 9,900 \\
8     & 0.1768  & ---     & 39,900 & 99,900 & 9,900 &  9,900 & 9,900 & 9,900 \\
12    & 0.1179 & 39,900 & 69,900 & 69,900 & 9,900 & 9,900 & 9,900 & 9,900 \\
16    & 0.08839 & 39,900 & ---     & ---     & ---    & ---     & --- & --- \\
20    & 0.07071 & 39,900 & ---     & ---     & ---    & ---     & --- & --- \\
\hline\hline
\end{tabular}
\end{center}
\end{table}

\begin{figure} 
\centerline{\includegraphics[width=0.8\textwidth]{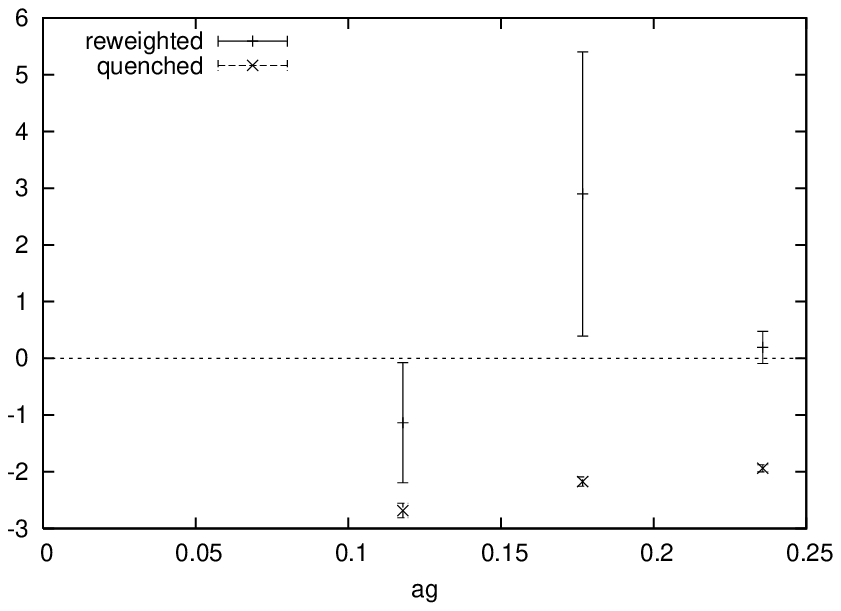}}
\caption{Real part of expectation values of the action density (over $g^2$)
with the periodic boundary condition. The parameters are identical to those of
entries in the $N_T/N_S=1$ column in Table~\ref{table:1}, except that the
number of configurations is 9,900 for each case.}
\label{fig:7}
\end{figure}

To give an idea of the quality of our numerical simulation and to illustrate
that the quantum effect of fermions is really taken into account, in
Fig.~\ref{fig:7}, we plot the real part of the expectation value of the
action density in Eq.~(\ref{fourxfifteen}) with the periodic boundary condition
as a function of the lattice spacing~$ag$.
Since the lattice action density is $Q$-exact, its expectation value under the
periodic boundary condition should be zero for any lattice spacing. The plot
is certainly consistent with this. On the other hand, the expectation values in
the quenched approximation are definitely not consistent with zero as they do
not contain the effect of dynamical fermions. See also Fig.~2 of
Ref.~\citen{Suzuki:2007jt}.

\begin{figure}
\centerline{\includegraphics[width=0.8\textwidth]{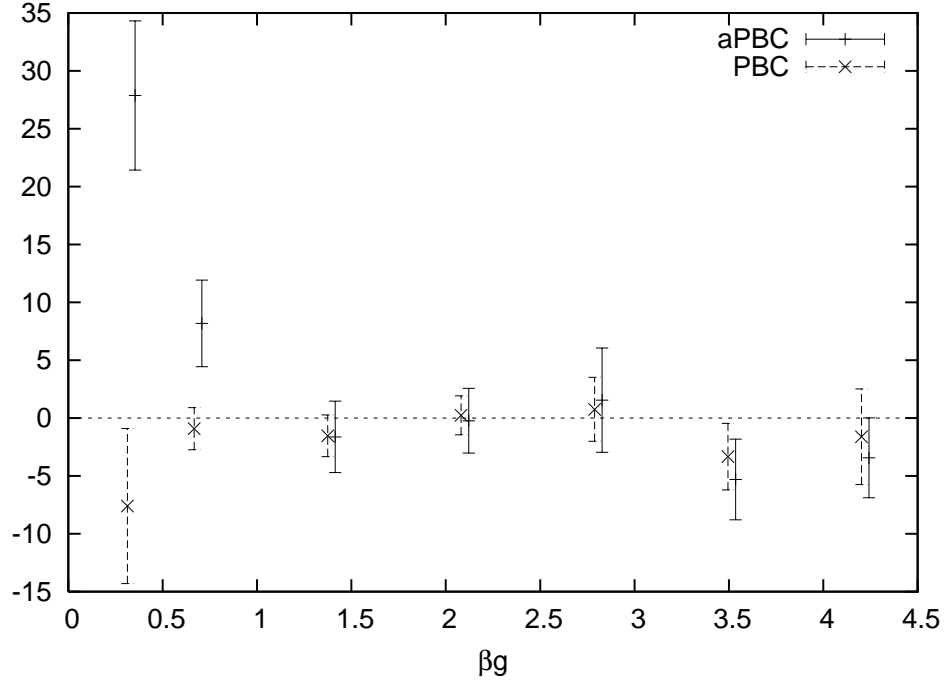}}
\caption{Continuum limit of the real part of the expectation value of the
Hamiltonian density,
$\lim_{a\to0}\Real\langle\mathcal{H}(x)\rangle_{\text{aPBC}}/g^2$
and $\lim_{a\to0}\Real\langle\mathcal{H}(x)\rangle_{\text{PBC}}/g^2$, as functions
of the temporal size~$\beta g$. The errors are only statistical ones. For the
periodic boundary condition, the number of configurations is 9,900 for all
cases.}
\label{fig:8}
\end{figure}

\begin{figure} 
\centerline{\includegraphics[width=0.8\textwidth]{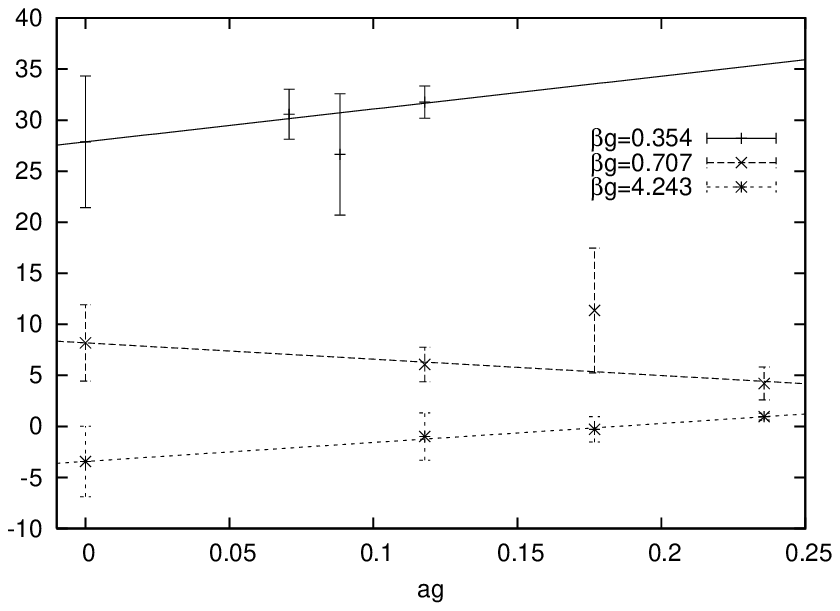}}
\caption{Linear extrapolations of
$\Real\langle\mathcal{H}(x)\rangle_{\text{aPBC}}/g^2$ to the continuum $a=0$ for
various values of $\beta g$. The errors are only statistical ones.}
\label{fig:9}
\end{figure}

Now, our main result in this paper is illustrated in Fig.~\ref{fig:8}. We
plotted the continuum limit of the real part of the expectation value of the
Hamiltonian density~(\ref{fourxfortythree}) with the antiperiodic boundary
condition, $\lim_{a\to0}\Real\langle\mathcal{H}(x)\rangle_{\text{aPBC}}$, as a
function of the physical temporal size of the system~$\beta g$.
For each $\beta g$, the continuum limit was obtained by a linear $\chi^2$-fit,
as depicted in Fig.~\ref{fig:9}.\footnote{The statistical errors in
Fig.~\ref{fig:9} are one standard deviation, obtained by jackknife analysis.
Jackknife analysis is necessary because we are using the reweighting
method, as explained in Ref.~\citen{Suzuki:2007jt}. The errors in the linear
$\chi^2$-extrapolation are estimated from the range of fitting parameters that
corresponds to a unit variation of $\chi^2$.}
For $\beta g\gtrsim1$, the expectation value rapidly approaches the asymptotic
value, that is, $\mathcal{E}_0/g^2$, according to Eq.~(\ref{fourxfortyfive}).
We may estimate the asymptotic value in $\beta g\to\infty$ by $\chi^2$-fit
using a constant. The use of four data points in Fig.~\ref{fig:8} at
$\beta g>2$ gives $\mathcal{E}_0/g^2={-}2.0\pm1.7$ and three points at
$\beta g>2.5$ gives $\mathcal{E}_0/g^2={-}3.0\pm2.2$.
If we use an exponential function $A\exp(-B\beta g)+C$ and all points in
Fig.~\ref{fig:8}, we have $\mathcal{E}_0/g^2={-}2.2\pm1.4$. All these results
on $\mathcal{E}_0/g^2$ are consistent and, at least within one standard
deviation, we do not observe positive vacuum energy density. We regard this as
an indication of the fact that supersymmetry is not dynamically broken in this
system. Of course, errors in our present result are large and we cannot exclude
the possibility of supersymmetry breaking of $O(1)$ in $\mathcal{E}_0/g^2$.
Further reduction of statistical errors will allow us to conclude whether the
scale of dynamical supersymmetry breaking is $O(1)$ or
not.\footnote{For the present lattice model, we are currently developing a
simulation code with the pseudo-fermion and the RHMC
algorithm.\cite{Clark:2004cp} \ We hope this will enable us to reduce the
statistical errors without substantially increasing the number of
configurations.}

In the present lattice model with the \emph{periodic\/} boundary condition, we
observed that the Pfaffian of the Dirac operator is almost real positive
(recall Fig.~\ref{fig:6}) and this implies that $\mathcal{Z}_{\text{PBC}}\neq0$
with finite lattice spacings (unlike the case of Fig.~\ref{fig:1}).\footnote{In
the quantum mechanical system~(\ref{threexthirtytwo}) in which supersymmetry is
spontaneously broken, the Witten index
$\mathcal{Z}_{\text{PBC}}$~(\ref{twoxtwo}) becomes zero because the fermion
determinant is not positive-definite, as shown in Fig.~\ref{fig:1}. In our
present lattice model for the two-dimensional $\mathcal{N}=(2,2)$ super
Yang-Mills theory, the Pfaffian is almost real positive, and thus,
$\mathcal{Z}_{\text{PBC}}\neq0$ with finite lattice spacings. One might then
think that this latter fact alone is sufficient to conclude that
supersymmetry is not spontaneously broken in this system. Although this
argument is not quite correct, because there is a possibility that the
coefficient $\mathcal{N}_{\text{PBC}}$, and thus $\mathcal{Z}_{\text{PBC}}$, in
Eq.~(\ref{twoxtwo}) becomes zero in the continuum limit, it certainly indicates
that dynamical supersymmetry breaking is unlikely in this system.} We can
thus consider the expectation values with the periodic boundary condition. In
Fig.~\ref{fig:8}, we have also plotted the real part of the expectation values
of the Hamiltonian density for various temporal sizes with the periodic
boundary condition. Since the Hamiltonian density is $Q$-exact, all the
expectation values with the periodic boundary condition must be zero, if one
can define them. The plot is clearly consistent with this. This also supports
the idea that supersymmetry is not broken in this system. If supersymmetry is
spontaneously broken, the expectation value
$\langle\mathcal{H}(x)\rangle_{\text{PBC}}$ must be indefinite, as
Eq.~(\ref{twoxfour}) shows. In our simulation, the expectation value is
computed as $\langle\mathcal{H}(x)\Pf\{D\}\rangle_{\text{quenched}}/
\langle\Pf\{D\}\rangle_{\text{quenched}}$,\cite{Suzuki:2007jt} and it can be
indefinite only when $\langle\Pf\{D\}\rangle_{\text{quenched}}=0$ in the
continuum limit. We did not see such a tendency and obtained the plot shown in
Fig.~\ref{fig:8}. Also, it is interesting to note that the overall feature of
Fig.~\ref{fig:8} is quite similar to that of Fig.~\ref{fig:5}, rather than that
of Fig.~\ref{fig:4}.

\section{Discussion}
The most direct way to observe the spontaneous supersymmetry breaking would be
to examine the degeneracy of boson and fermion mass spectra through
two-point correlation functions. Although this method is conceptually clear, a
reliable exponential fit of two-point functions would require a rather large
lattice extent. The method we propose in this paper is computationally much
easier because it is based on the measurement of one-point functions, the
expectation values of a Hamiltonian (density). A weakness is the ambiguity in
the choice of the Hamiltonian in the Euclidean lattice formulation. In this
paper, we gave a justification (on the basis of a topological property of the
Witten index) of the choice, for lattice formulations that possess a manifestly
preserved fermionic symmetry~$Q$.
In any case, this is the first work of a direct investigation of the
spontaneous supersymmetry breaking in a gauge field model (for which the Witten
index is unknown) by numerical simulation. Before the recent developments in
the lattice formulation of supersymmetric gauge
theories,\cite{Kaplan:2003uh,Feo:2004kx,Giedt:2006pd,Giedt:2007hz}
one could not even imagine such a study feasible.

One may ask the extent of the applicability of our method. We already
have a lattice formulation\cite{Sugino:2004qd} of the two-dimensional
$\mathcal{N}=(4,4)$ super Yang-Mills theory in which two fermionic symmetries
are exactly preserved. Thus, it should be possible to study possible
dynamical supersymmetry breaking in this theory in a similar manner. For
other supersymmetric field theories (except Wess-Zumino-type models), strictly
speaking, we do not have a lattice formulation with \emph{manifestly\/}
preserved fermionic symmetry. Further study is needed on the lattice
formulation of these theories (including physically interesting models, such as
two-dimensional supersymmetric nonlinear sigma models, three-dimensional
supersymmetric pure Yang-Mills theories, and supersymmetric gauge theories with
matter multiplets). For related works, see
Refs.~\citen{Catterall:2003uf,Catterall:2006sj,Sugino:2004uv,Endres:2006ic}.

Suppose that we have a lattice formulation in which the lattice action~$S$ and
the integration measure~$d\mu$ are manifestly invariant under a fermionic
transformation~$Q$. It is quite conceivable that, for typical models in
dimensions higher than two, this manifest $Q$-invariance alone is not
sufficient to ensure automatic restoration of the invariance under a full set
of supersymmetry transformations. One would then have to supplement a
counter term $\Delta S$ to the original lattice action. However, it is also
conceivable that $\Delta S$ is invariant under $Q$, because the original
lattice regularization preserves a manifest $Q$-invariance. If this is true, it
is again natural to adopt the prescription for the Hamiltonian
$H\equiv iQ\overline{\mathcal{Q}}/2$ because the relation
\begin{equation}
   \int_{\text{PBC}}d\mu\,He^{-S-\Delta S}
   =\int_{\text{PBC}}d\mu\,\frac{i}{2}Q\overline{\mathcal{Q}}\,e^{-S-\Delta S}
   =\int_{\text{PBC}}d\mu\,Q
   \left(\frac{i}{2}\overline{\mathcal{Q}}\,e^{-S-\Delta S}
   \right)
   =0,
\end{equation}
which corresponds to the topological property of the Witten index, still holds.

Can we not do anything if the $Q$-invariance is not manifest in the lattice
formulation that is adopted? A natural idea is to take an arbitrarily
chosen Hamiltonian $\widetilde H$ and then subtract a constant from it,
$H=\widetilde H-c$, such that relation~(\ref{twoxthree}) holds. It is easy
to see that this requirement implies\footnote{Incidentally, this formula
reproduces a prescription for the origin of the energy in
Refs.~\citen{Catterall:2007fp} and \citen{Anagnostopoulos:2007fw} in which the
thermal average of the energy in one-dimensional supersymmetric Yang-Mills
theories is numerically studied. The thermal average of the energy is given by
(the minus) the $\beta$-derivative of the thermal partition
function~(\ref{twoxfive}). In one-dimensional supersymmetric Yang-Mills
theories, by rescaling the imaginary time and dynamical variables, one sees
that a Hamiltonian~$\widetilde H$ is simply given by $-3/\beta$ times the
Euclidean \emph{action\/} up to an additive constant. If one substitutes this
$\widetilde H$ into Eq.~(\ref{twoxten}), one ends up with the formulas in
Refs.~\citen{Catterall:2007fp} and~\citen{Anagnostopoulos:2007fw}. Assuming
that $\int_{\text{PBC}}d\mu\,e^{-S}\neq0$, the second term in
Eq.~(\ref{twoxten}) can be evaluated by the lowest-order perturbation theory.}
\begin{equation}
   H=\widetilde H
   -\frac{\int_{\text{PBC}}d\mu\,\widetilde He^{-S}}
   {\int_{\text{PBC}}d\mu\,e^{-S}}.
\label{twoxten}
\end{equation}
However, \emph{when} supersymmetry is spontaneously broken, the denominator of
the second term of Eq.~(\ref{twoxten}) would vanish (because it is proportional
to the Witten index) and thus, unfortunately, it appears that
formula~(\ref{twoxten}) itself cannot be used for cases in which supersymmetry
is spontaneously broken. We certainly need a more elaborate idea.

Our main aim in this work was to examine a possible spontaneous
supersymmetry breaking from the ground-state (vacuum) energy obtained by
$\lim_{\beta\to\infty}\langle H\rangle_{\text{aPBC}}$. It is nevertheless
important to study $\langle H\rangle_{\text{aPBC}}$ with \emph{finite\/}
$\beta$ because it contains useful information on the energy spectrum of
\emph{excited states}. That is, when there is an energy gap $\Delta E$ between
the first excited state and the ground-state, the decay of
$\langle H\rangle_{\text{aPBC}}$ for $\beta\to\infty$ is exponential,
$\sim\exp(-\beta\Delta E)$, whereas when the spectrum is continuous starting at
zero and the density of states behaves as $\rho(E)\sim E^{\nu-1}$, the decay of
$\langle H\rangle_{\text{aPBC}}$ for $\beta\to\infty$ is power-like,
$\sim\nu/\beta$.

The behavior in Figs.~\ref{fig:4} and~\ref{fig:5} appears to be consistent
with the exponential decay expected for quantum mechanical systems
with discrete spectra. For the two-dimensional $\mathcal{N}=(2,2)$ super
Yang-Mills theory in~Fig.~\ref{fig:8}, is the decay an exponential or power
one? The error bars in the figure are too large for a reliable fit and we
reserve this study for a future work. If the decay turns out to be exponential,
it will be very intriguing because it will imply that an energy gap opens up
owing to interactions. Note that a weak coupling analysis shows that the
spectrum is continuous starting at zero
\emph{even in a finite volume\/} because of noncompact flat directions of the
classical potential energy.\footnote{As shown in Ref.~\citen{Fukaya:2006mg}, to
all orders of perturbation theory, there exists a massless bosonic state in
this system with infinite volume. Even if this persists nonperturbatively, the
state cannot produce the spectrum that is continuous starting at zero because
the spatial momentum is discrete with a finite volume.} Then this system
provides an example in which the Witten index becomes well defined as a result
of interactions while a perturbative analysis indicates that it is not so.

\section*{Acknowledgements}
I.~K.\ would like to thank Makiko Nio for a useful comment.
F.~S.\ would like to thank Kentaro Hori for e-mail correspondence, and the
Niels Bohr Institute for their hospitality in the final stage of this work.
H.~S.\ would like to thank Ko Furuta, Masanori Hanada and Tomohisa Takimi for
discussions. The results for the two-dimensional model were obtained using
the RIKEN Super Combined Cluster (RSCC). I.~K.\ is supported by the Special
Postdoctoral Researchers Program at RIKEN. The work of H.~S.\ is supported in
part by a Grant-in-Aid for Scientific Research, 18540305, and by JSPS and
French Ministry of Foreign Affairs under the Japan-France Integrated Action
Program (SAKURA).

\end{document}